\documentclass[aps,preprintnumbers,prd,twocolumn,superscriptaddress,nofootinbib]{revtex4-1}
\usepackage{blindtext}
\usepackage{lineno}
\usepackage{bm,color,xcolor}
\usepackage{slashed} 
\usepackage{graphicx}
\usepackage{subfigure} 
\usepackage{soul} 
\usepackage{amsmath}
\usepackage{multirow}
\usepackage{float}
\usepackage{comment}
\usepackage{fontawesome} 
\usepackage{mathtools}
\usepackage[colorlinks=true
,urlcolor=blue
,anchorcolor=blue
,citecolor=blue 
,filecolor=blue
,linkcolor=blue
,menucolor=blue
,linktocpage=true
,pdfproducer=medialab
,pdfa=true
]{hyperref}
\usepackage[nameinlink,capitalize]{cleveref} 
\usepackage{lipsum}

\makeatletter
\def\l@subsubsection#1#2{}
\makeatother

\newcommand{\beq}{\begin{equation}}
\newcommand{\eeq}{\end{equation}}
\newcommand{\bea}{\begin{eqnarray}}
\newcommand{\eea}{\end{eqnarray}}

\usepackage[arrows]{ezedits}

\begin{document}


\preprint{UMN-TH-4524/26}

\title{A Busy Higgs Signal}

\author{Peiran Li} 
\email{li001800@umn.edu}
\author{Zhen Liu}
\email{zliuphys@umn.edu}
\affiliation{School of Physics and Astronomy, University of Minnesota, Minneapolis, MN 55455, USA}
\author{Lian-Tao Wang} 
\email{liantaow@uchicago.edu}
\affiliation{Department of Physics, University of Chicago, Chicago, IL, 60637, USA}
\affiliation{Enrico Fermi Institute, University of Chicago, Chicago, IL, 60637, USA}
\affiliation{Kavli Institute for Cosmological Physics, University of Chicago, Chicago, IL 60637, USA}
\affiliation{Leinweber Institute for Theoretical Physics, University of Chicago, Chicago, IL 60637, USA}
\affiliation{HEP Division, Argonne National Laboratory, 9700 Cass Ave., Argonne, IL 60439, USA}

\begin{abstract}
Higgs final states are prime  targets in the search for physics beyond the Standard Model. In the conventional picture, $SU(2)$ symmetry together with the Goldstone Equivalence Theorem correlates Higgs and gauge-boson final states, implying comparable sensitivity in channels such as 
$hh$, $ZZ$, and $WW$ in searches for heavy resonances. In this work, we identify a mechanism to parametrically violate this expectation. We show that higher-order Higgs couplings can induce an electroweak-symmetry-breaking enhancement that selectively amplifies Higgs-rich final states, allowing them to become the leading discovery channels of new resonances. For scalar resonances, this can make di-Higgs the dominant bosonic signal. 
For resonance masses higher than a couple of TeV, it also opens resonant tri-Higgs and four-Higgs channels as well-motivated search targets. The same underlying mechanism extends to heavy fermionic and vector resonances, where it can similarly enhance channels such as $ht$, $Zh$, and $\gamma h$. We present this framework in effective field theory, demonstrate possible UV completions, and discuss its implications for collider searches.
\end{abstract}

\maketitle

\setcounter{secnumdepth}{3}
\setcounter{tocdepth}{1}

\section{Introduction}
\label{sec:introduction}

Since the discovery of the Higgs boson~\cite{ATLAS:2012yve,CMS:2012qbp}, the Higgs sector has been studied extensively, with increasing attention devoted to precision measurements of Higgs couplings~\cite{ATLAS:2022vkf,CMS:2022dwd,Ruan:2021gap,An:2018dwb,Tian:2013yda,Barklow:2017suo,dEnterria:2017dac,Maura:2025rcv,deBlas:2022aow,Forslund:2022xjq,Forslund:2023reu,Li:2024joa}. 
Motivated by accessing the Higgs self interaction, the di-Higgs final state is one of the main focuses of the physics program at the LHC~\cite{DiVita:2017vrr,Cepeda:2019klc,DiMicco:2019ngk,Maltoni:2024dpn,CMS:2026nuu}. In parallel, substantial effort has also been devoted to searches for resonances in the di-Higgs channel~\cite{CMS:2024phk,ATLAS:2023vdy,CMS:2026nuu}. 

As we will review in this paper, in general, $SU(2)$ invariance and the Goldstone Equivalence Theorem dictate relations between rates of the di-Higgs and other di-gauge boson final states, such as $ZZ$ and $W^+ W^-$, in particular for resonances much heavier than the weak scale. In this case, we would expect that the sensitivities in all of these channels would be broadly comparable~\cite{ATLAS:2023vdy,ATLAS:2020fry,CMS:2019kaf}. 

In this work, we highlight a scenario in which the Higgs channel is qualitatively distinct. We call this \textit{the busy Higgs mechanism}, for which the di-Higgs can be the leading discovery channel. Similar enhancement relative to the gauge boson channels can also be achieved in the single Higgs final states, such as $ht$ and $hZ$ in new fermionic or vector resonances, respectively.

We begin with the case of a spin-0 resonance, $S$, with a coupling through the Higgs portal
\begin{equation}
S \, H^\dagger H.
\label{coupling_basic}
\end{equation}

In our discussion, we take
\begin{align}
H = \frac{1}{\sqrt{2}}
\begin{pmatrix}
a_1 + i a_{2} \\
(v + h) + i a_{0}
\end{pmatrix},
\end{align}
where $a_i$ denotes the Goldstone modes. An equivalent derivation can also be given in unitary gauge which is discussed in \cref{sec:br}. The Higgs portal operator can then be written as
\begin{align}
H^\dagger H
= \frac12 \left( (v + h)^{2} + \sum_{i=0}^{2} a_{i}^{2} \right).
\end{align}
At the quadratic order in the fields, this operator displays a symmetry between $a_i$ and $h$. This structure is a direct consequence of the $SU(2)$ symmetry since the Higgs Vacuum Expectation Value (VEV) does not enter at this order. By contrast, terms proportional to the VEV break this symmetry, and we will investigate their effects. 
Hence, for the coupling in \cref{coupling_basic}, the decay branching ratios (BR) of $S$ satisfy $\text{BR}(S\to hh) = \text{BR}(S \to a_i a_i)$. For $m_S \gg v$, we have 
\begin{equation}
    \text{BR}(S\to h h) = \text{BR}(S \to Z Z) = \frac{1}{2} \text{BR}(S \to W^+ W^-). 
\end{equation}
The same pattern persists for other Higgs bilinear operators, such as $\partial_\mu S \partial^\mu (H^\dagger H)$ and $S|D_\mu H|^2$. These operators can equivalently be viewed as inducing $S$-$h$ mass mixing, $S$-$h$ kinetic mixing, and direct couplings of $S$ to Higgs and SM gauge bosons. All of them maintain such ratios between decay channels with $O(v/m_S)$ corrections. 
Hence, we expect di-Higgs resonance searches to have discovery power comparable to that of the 
$ZZ$ and 
$WW$ channels, up to differences in signal–background discrimination at a given collider experiment. One can further show that the same conclusion extends to heavy vectors and spin-2 resonances, with an analogous relation for heavy fermions as well~\cite{Aguilar-Saavedra:2013qpa,DeSimone:2012fs,Xiao:2014kba,Pappadopulo:2014qza,Antipin:2007pi,Agashe:2007zd}. This leads to the central question of this paper: is it impossible to have Higgs final states driving the discovery?\footnote{Near the low mass regime where $m_S\sim v$, one can have sizable enhancements or suppressions~\cite{Dolan:2012ac,No:2013wsa,Chen:2014ask,Kotwal:2016tex,Lewis:2017dme}.}

\section{The busy Higgs mechanism}

{\flushleft \textbf{Busy Scalar}:} Contrary to this common lore, we show there is a new class of theories where Higgs final states will be the driving discovery mode! We first consider coupling $S(H^\dagger H)^n$, which is
\begin{align}
{\cal L}  & \supset S(H^\dagger H)^n \label{eq:busyscalar_coupling} \\
& \to \frac{(v^2)^{n-1}}{2^n}S \left(\Sigma_i  a_i^2 \binom{n}{1}+  h^2 \binom{n}{1}+   4h^{2} \binom{n}{2} + \dots\right). \nonumber
\end{align}
The first two terms still preserve the $SU(2)$ relation between $a_i$ and $h$. However, the third term gives rise to di-Higgs coupling proportional to \(n^{2} \). This is an effect of Electroweak Symmetry Breaking (EWSB). What's remarkable is that this can happen even with $m_S \gg v$. The partial width of these di-boson decay channels vanishes as $v\to 0$. However, the ratio
\begin{align}
\frac{\text{BR}(S\to hh)}{\text{BR}(S\to a_i a_i)} = (2n-1)^2 
\end{align}
stays as a constant even with $v/m_S \to 0$. Therefore, unlike the usual expectation from $SU(2)$ symmetry, the di-Higgs channel can remain parametrically enhanced relative to the di-gauge boson channels, and could become the dominant discovery mode.

Multi-Higgs final states can also become relevant at sufficiently large $m_S$. While three-body decays are phase-space suppressed relative to two-body decays, 
this suppression becomes less severe as $m_S$ increases, since the integrated 
three-body phase space itself grows with $m_S^2$. Parametrically, the $n$-body phase-space factor
\begin{align}
\Phi_n=\int\delta^4(P-\sum^n_i p_i)\prod^n_i \frac{d^3 p_i}{(2\pi)^3 2E_i}
\end{align}
carries a positive mass dimension for $n>2$. In particular, the two-body factor scales as $1/(8\pi)$, whereas the three-body phase space scales as $m_S^2/(256\pi^3)$~\cite{Kumar:1969jjy}. In addition, the $Sh^3$ interaction from the busy operator receives a combinatorial enhancement, scaling as $\sim 2n(2n-1)(2n-2)/3!$. Taken together, these effects make a multi-Higgs-dominated decay pattern possible.

\begin{figure}
    \centering
    \includegraphics[width=0.85\linewidth]{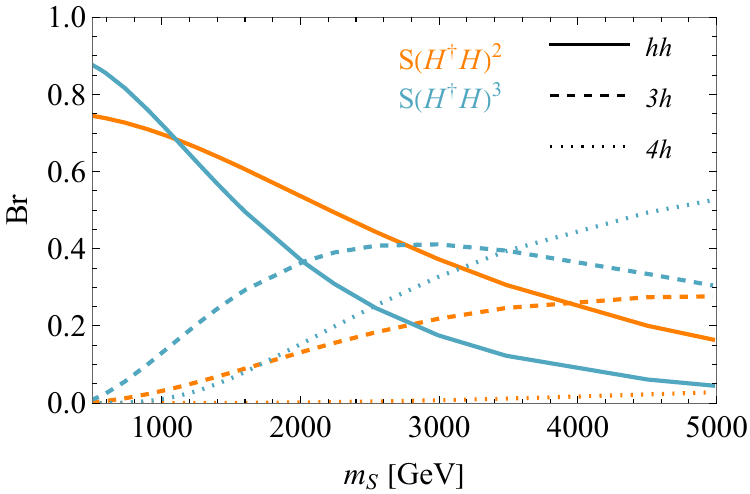}
    \caption{BR of $S$ to multi-Higgs as a function of $m_S$ based on different operators.  
    For higher dimensional operators, the multi-boson channels will be further enhanced (see \cref{sec:br}). 
    }
    \label{fig:Br_hh_hhh}
\end{figure}

For example, $S(H^\dagger H)^n$ will provide
\begin{align}
\frac{\Gamma(S\to hhh)}{\Gamma(S\to hh)}\sim \frac{1}{(4\pi)^2}\frac{2(n-1)^2}{3}\frac{m_S^2}{v^2}.
\end{align}
As shown in \cref{fig:Br_hh_hhh}, for sufficiently large $m_S$, the multi-Higgs channel could dominate. Multi-gauge-boson final states can also receive combinatorial enhancements, but only when they arise from higher multiplicity structure. 
For example, the combinatorial enhancement of the 
$hZZ$ or $hWW$ decay follows the same scaling as di-Higgs decay, rather than tri-Higgs decay (see \cref{sec:br}). Overall, pure-Higgs final states remain particularly attractive discovery channels, especially for large $n$. 

{\flushleft \textbf{Higgs-infused Fermion}:} This busy Higgs mechanism can be generalized in a straightforward manner to heavy fermionic resonances. Consider a heavy $SU(2)$ singlet fermion $T$ with coupling
\begin{align}
{\mathcal L} \supset \bar{q} \tilde{H}T + \text{h.c.} ,
\end{align}
where $q=(t, b)_L$ is a SM doublet fermion. From this interaction, we can derive the relations
\begin{align}
\text{BR}(T \to Z_L t) = \text{BR}(T \to  ht) = \frac{1}{2}\text{BR}(T \to W_L b). 
\end{align}
However, if the coupling instead takes the form
\begin{align}
{\mathcal L} \supset \frac{c_n}{\Lambda^{2n}}(H^\dagger H)^n \bar{q} \tilde{H}T + \text{h.c.} ,
\end{align}
we will have an $O(n^2)$ enhancement to the $T\to ht$ channel. 

{\flushleft \textbf{Higgs-infused Vector}:}  The mechanism can be generalized to vector resonances as well. 
A simple realization arises by introducing kinetic mixing with additional heavy vector field $Z'$:
\begin{align}
{\cal L} \supset\frac{1}{2}\frac{c_n}{\Lambda^{2n}}Z'_{\mu \nu} B^{\mu \nu} (H^\dagger H)^n, \label{eq:busy_vector}
\end{align}
where $B$ is the SM Hypercharge field. After the EWSB and properly taking into account the kinetic mixing proportional to $\epsilon = c_n\left(\frac{v^2}{2\Lambda^2}\right)^n$, this leads to 
\begin{equation}
{\cal L} \supset \binom{n}{1} \epsilon Z'_{\mu \nu} B^{\mu \nu}  \frac{h}{v} + \epsilon Z'_{\mu} J_\text{SM}^{\mu} + {\cal O}\left(\epsilon \frac{m_Z^2}{m_Z'^2}\right),
\end{equation}
where $B=c_W\gamma-s_W Z$ and $J_\text{SM}$ includes fermion current and Goldstone current. If only the second term were present, it would imply
\begin{align}
    \text{BR}(Z' \to Zh) = \text{BR}(Z' \to W^+ W^-).
\end{align}
The first term, however, yields an $n^2$ enhancement of the $Z' \to Zh$ and $Z' \to \gamma h$ amplitudes. Decays to a given SM fermion pair and to $WW$ are parametrically suppressed relative to $Zh/\gamma h$, since the $Z'_{\mu\nu}B^{\mu\nu}h$ interaction is momentum enhanced, whereas the $Z'_\mu J_\text{SM}^\mu$ couplings are not. 
For example, for a $Z^\prime$ with $m_{Z'}=1~\mathrm{TeV}$ and $n=1$, we find
\begin{align}
    \frac{\Gamma(Z'\to Zh~\text{and}~\gamma h)}{\Gamma(Z'\to WW)}\approx30~.\label{eq:Z_decay_enhancement}
\end{align}
Therefore, $\text{BR}(Z'\to Z/\gamma + h)$ provides a significant bosonic discovery channel.

{\flushleft \textbf{Higgs-infused Tensor}:} There can be also generalizations of the busy Higgs to the spin-2 resonances. The (minimal) spin-2 coupling to the Higgs doublets can be written as 
\begin{align}
{\cal L} \supset h_{\mu\nu} D^{\mu} H^{\dagger} D^{\nu} H.
\end{align}
This preserves the $SU(2)$ relation
\begin{align}
\text{BR}(h_{\mu\nu} \to hh) & = \text{BR}(h_{\mu\nu} \to Z_L Z_L) \nonumber \\ & = \frac{1}{2} \text{BR}(h_{\mu\nu} \to W_L W_L). 
\end{align}
Multiplying the above interaction by $(H^\dagger H)^n$ will generate enhanced multi-Higgs ($n_h>2$) decay channels, although this would not change the two-body BR relations since the minimal coupling already saturates the di-boson final state.

\section{UV completions}

To ensure the dominance of the busy Higgs mechanism, we want the operator with $n>1$ to be the leading one. Generating such an operator at leading order is not difficult to imagine, in particular, through some notion of locality in theory space. For example, if we have the following interactions
\begin{align}
\Phi N_1 N^\dagger_{n+1}, \ N_1 N^\dagger_2 H^\dagger H, \ ... \ , \  N_n N^\dagger_{n+1} H^\dagger H,
\end{align}
where $\Phi,N_i$ are additional scalars. Then, the coupling in \cref{eq:busyscalar_coupling}
can be generated at 1-loop with $S=\text{Re}(\Phi)$. However, since $H^\dagger H$ is a singlet, there is no symmetry that forbids lower-dimensional operators at tree level, such as $SH^\dagger H$.
One possibility is to use the operator $H_u\epsilon H_d$ in a two-Higgs doublet setup, with additional global charges assigned. Hence, we will be able to control the leading order at which this operator appears in the theory. 
Parameterizing $H_u=(v_u+H_u^0, \sqrt{2}H_u^-)/\sqrt{2}$ and $H_d = (\sqrt{2}H_d^+, v_d + H_d^0)/\sqrt{2} $, in the Higgs basis, we have
\begin{align}
H_u \epsilon H_d 
&\supset\frac{1}{2}(-v \sin \beta \sin \alpha + v \cos \beta \cos \alpha)h  -\frac{\sin 2\alpha}{4} h^2 \nonumber \\ &
 ~\quad + \frac{\sin 2\beta}{4} a_0^2 + \frac{\sin 2\beta}{2}a^+ a^- + ...,\nonumber\\
&\simeq \frac{\sin 2 \beta }{4} \left( 2 vh +  (h^2 + a_0^2 +2a^+ a^-)  \right) + ...
\end{align}
where $\tan \beta = v_u/v_d$, and we have taken the decoupling limit, $\sin \alpha \simeq -\cos \beta$ and $\cos \alpha \simeq \sin \beta$. The ``$+ ...$'' contains the heavy doublet without a VEV and misalignment corrections. If we look at the terms quadratic in fields, we again see a $SU(2)$ symmetry between the light Higgs and the Goldstones. Starting with an operator $S(H_u\epsilon H_d)^n$, we will obtain an effective operator $S(H^\dagger H)^n$ after integrating out the heavy scalars. 
In \cref{sec:UV_model}, we explicitly show a couple of examples in which the busy Higgs operator can be generated.

Starting from an operator $S(H^\dagger H)^n$, one can always close the loop with one pair of $H^\dagger H$ and obtain another operator with a lower $n$. Of course, this lower-dimensional operator is suppressed by a loop factor $(4 \pi)^{-2}$. This expectation can be explicitly verified by computing threshold corrections in a UV completion, as we present in \cref{sec:UV_model}.

\section{Busy Higgs at colliders}
\begin{figure}
    \centering
    \includegraphics[width=0.32\linewidth]{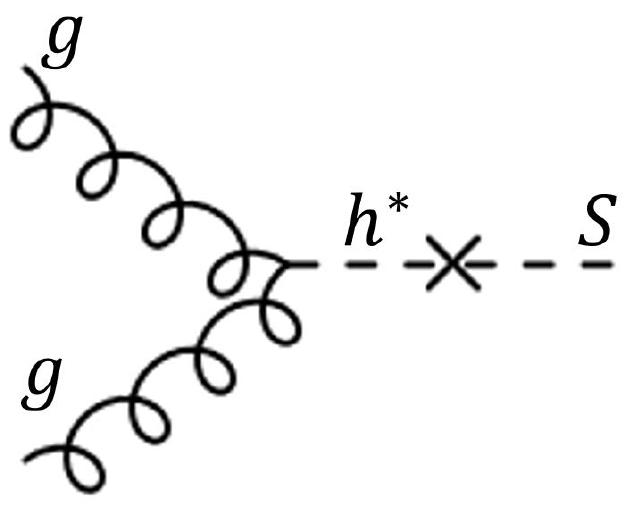}~~~~
    \includegraphics[width=0.3\linewidth]{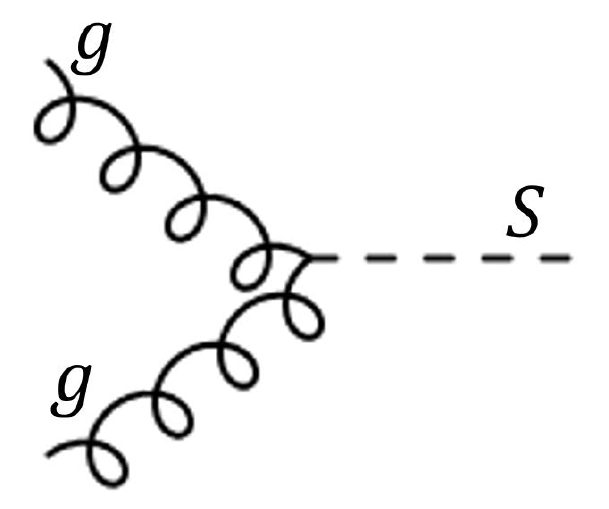}
    \caption{\textbf{Left:} The main production channel of heavy scalar $S$ at LHC via mass-mixing with Higgs. The VBF production can be important in the high-mass regime. \textbf{Right:} The main production channel of heavy scalar $S$ at LHC via $SG^{\mu\nu}G_{\mu\nu}$.
    }
    \label{fig:production_channel}
\end{figure}

As analyzed in the previous sections, the busy scalar motivates the di-Higgs channel with a dominant $S\to hh$ branching ratio. At the LHC, it is produced predominantly via gluon fusion, either through $S$-$h$ mixing or through a direct coupling to gluons, as shown in \cref{fig:production_channel}. For a simple and direct interpretation and comparison between different busy Higgs operators, we use the direct coupling from the effective Lagrangian
\begin{align}
c_{s}\frac{\alpha_s}{12\pi \Lambda} S G^a_{\mu\nu}G^{a\mu\nu},
\end{align}
which can arise from loop effects of any vector-like heavy colored fermion and coupled with $S$.

This operator also opens up $S\to gg$ and modifies the branching ratio:
\begin{align}
&\frac{\text{BR}(S\to gg)}{\text{BR}(S\to hh)}\\
=& \frac{8\times4^n}{[2n(2n-1)]^2}\left(\frac{c_S}{c_n}\right)^2
\left(\frac{\alpha_s}{12\pi}\right)^2
\left(\frac{\Lambda}{v}\right)^{4n-4}
\left(\frac{m_S}{\Lambda}\right)^4
 .\nonumber
\end{align}
Taking a TeV scale benchmark choice $(\Lambda=m_S=1~\text{TeV},~c_S=c_n=1,~n=2)$, the di-gluon decay width is smaller than di-Higgs decay width:
\begin{align}
\frac{\text{BR}(S\to gg)}{\text{BR}(S\to hh)}=0.2\%~.
\end{align}
More generally, the relative size of $S\to gg$ and $S\to hh$ is UV-model dependent and can vary significantly across different benchmarks.~\footnote{In our collider analysis, we set $\text{BR}(S\to gg)\approx 0$ for simplicity; scenarios with a sizable gluonic branching fraction can be recovered by a simple rescaling of the signal rate.
In particular, if the busy operator also originates from a loop, $\text{BR}(S\to gg)$ can be significant.}

\begin{figure}
    \centering
    \includegraphics[width=0.95\linewidth]{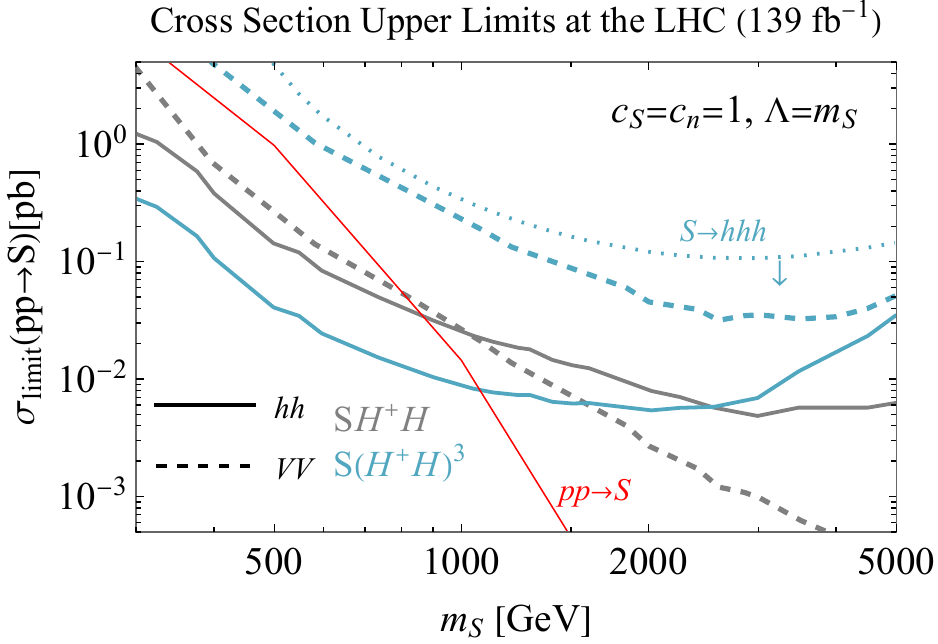}
    \caption{Existing 95\% C.L. upper limits on the heavy-scalar production cross section from different channels for two benchmark operators. Solid/Dashed lines are from di-Higgs/di-gauge-boson searches~\cite{ATLAS:2023vdy,ATLAS:2020fry}. The dotted curve is derived from the inclusive SM tri-Higgs search~\cite{CMS-PAS-HIG-24-012}; a resonance-based tri-Higgs analysis would substantially improve the sensitivity. The red curve refers to the production cross section via $SG^{\mu\nu}G_{\mu\nu}$. 
    }
    \label{fig:busy_scalar_projection_xs}
\end{figure}

We now present existing limits and projected reaches at the LHC. 
\cref{fig:busy_scalar_projection_xs} shows the current 95\% C.L. upper bounds on the heavy-scalar production cross section, where the solid/dashed/dotted curves are obtained from di-Higgs/di-gauge-boson/tri-Higgs searches~\cite{ATLAS:2023vdy,CMS:2024phk,ATLAS:2020fry,ATLAS:2019nat,ATLAS:2017tlw,CMS:2022pjv,CMS:2018dff,CMS-PAS-HIG-24-012,ATLAS:2024xcs}. 
For each busy Higgs benchmark, we rescale the corresponding experimental limits by the operator-dependent branching ratios to derive bounds on $\sigma(pp\to S)$. 
The red curve, on the other hand, shows the production cross section for $pp\to S$ induced by the $SG^{\mu\nu}G_{\mu\nu}$ interaction, with $c_S=c_n=1$ and $\Lambda=m_S$.

As expected from the enhancement of the decay mode $S\to hh$ in the busy Higgs mechanism, di-Higgs searches are significantly more sensitive than the di-gauge boson channels, as indicated by the blue solid curve, which lies below the blue dashed curve. The dotted curve is derived from the inclusive SM tri-Higgs limit~\cite{CMS-PAS-HIG-24-012} rather than a dedicated resonant $S\to hhh$ search; a resonance-based tri-Higgs analysis would therefore be expected to substantially improve the sensitivity with low background. 
Finally, for larger $n$ (e.g.\ $n\geq 3$), one may also anticipate enhanced multi-boson signatures, such as $hVV$ and $hhhh$. In particular, purely multi-Higgs final states can receive additional combinatorial enhancements at large $n$, leading to larger branching fractions.

\begin{figure}
    \centering
    \includegraphics[width=0.95\linewidth]{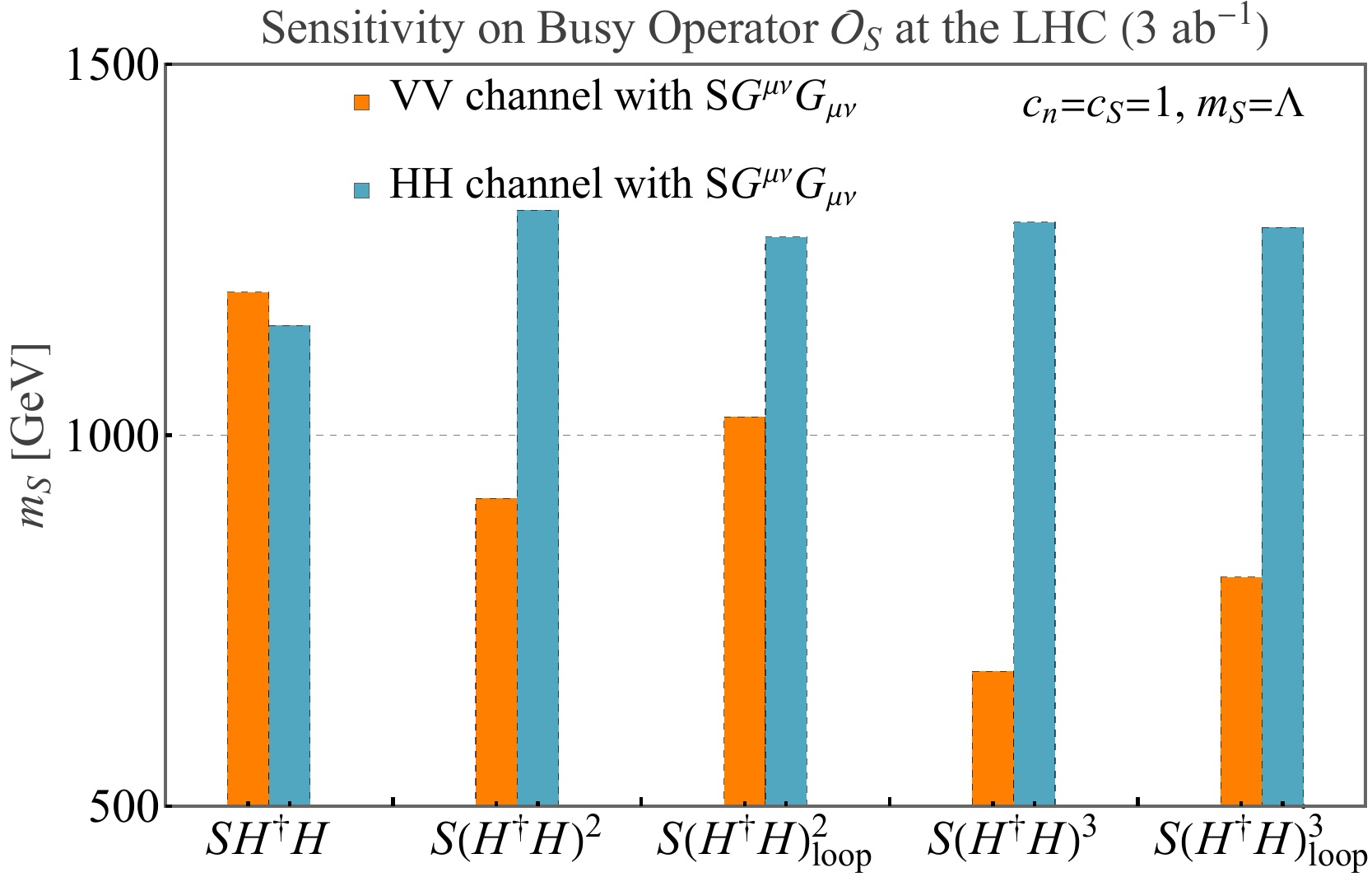}
    \caption{Projected 95\% C.L. mass reaches of heavy scalar at the LHC, comparing the di-gauge boson and di-Higgs channels (rescaled from existing searches~\cite{ATLAS:2023vdy,ATLAS:2020fry}). The first operator is treated as $\frac{m_S^2}{\Lambda}SH^\dagger H$. The heavy scalar $S$ is assumed to be produced via gluon fusion through $SG^{\mu\nu}G_{\mu\nu}$. For $S(H^\dagger H)^n_\text{loop}$, we include the impact of loop-induced lower-dimensional operators.
    }
    \label{fig:busy_scalar_projection_op}
\end{figure}

\cref{fig:busy_scalar_projection_op} then translates these sensitivities into mass reaches on the busy operators for the HL-LHC with an integrated luminosity of $3~\text{ab}^{-1}$. In particular, the di-Higgs channel probes the busy operators up to higher mass reach than other channels. Including higher-order effects in $S(H^\dagger H)^n_\text{loop}$, loop corrections can induce lower-dimensional operators, but the projected reach changes only mildly. As discussed in \cref{sec:loop_eft}, the corresponding correction to the decay amplitude scales as
\begin{align}
\mathcal{M}_\text{loop}\sim\mathcal{M}_\text{tree}\left(1+\frac{n^2}{16\pi^2}\frac{\Lambda^2}{v^2}\right),
\end{align}
and remains subdominant to the tree-level contribution for $\Lambda\lesssim 4\pi v/n$. We also note that there could be additional suppression depending on the UV completion. In this sense, our estimate should be regarded as a conservative assessment of the size of the loop contribution.

\begin{figure}
    \centering
    \includegraphics[width=0.95\linewidth]{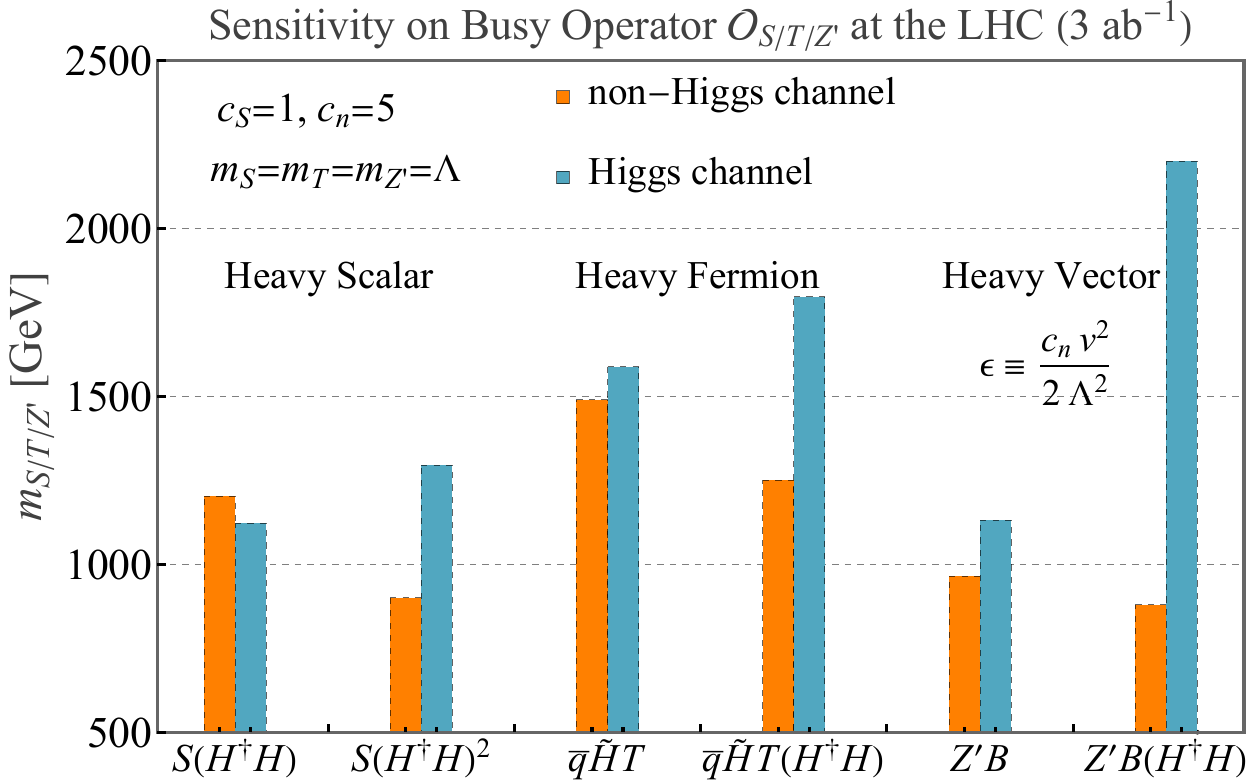}
    \caption{
    Projected 95\% C.L. mass reach at the LHC for heavy resonances in the minimal and busy-operator scenarios. The orange and blue bars denote the most sensitive non-Higgs and Higgs final states, respectively. The heavy-fermion reaches are obtained by rescaling existing searches in~\cite{CMS:2022fck}, while the heavy-vector reaches are obtained from~\cite{ATLAS:2020fry,ATLAS:2022enb,ATLAS:2020jeb}. We adopt the benchmark choice $c_n=5$ and $m_T=m_{Z'}=\Lambda$; for the vector case, the kinetic mixing is always parameterized by $\epsilon \equiv c_n v^2/(2\Lambda^2)$.
    }
    \label{fig:busy_fermion_projection_op}
\end{figure}

Furthermore, the heavy-fermion and heavy-vector cases are summarized in \cref{fig:busy_fermion_projection_op,}, where we present the projected 95\% C.L. mass reaches based on various searches at the LHC~\cite{CMS:2022fck,ATLAS:2018ziw,ATLAS:2020fry,ATLAS:2022enb,ATLAS:2020jeb,CMS:2024hbn,CMS:2022pjv,CMS:2025rqr}. For the heavy fermion $T$, the reach is determined primarily by the $gg\to T\bar{T}$ production rate together with the decay branching ratios. The main qualitative effect already appears at the $n=1$ level. In the minimal case ($n=0$), ${\rm BR}(T\to ht)\simeq 1/4$, so the Higgs channel is not dominant. Once the busy operator $\bar{q}\tilde{H}T(H^\dagger H)$ is introduced, however, the $ht$ branching fraction becomes large, enhancing the Higgs-channel sensitivity. Higher-order busy operators may further enhance the Higgs-channel sensitivity, although only modestly, since the $T\to ht$ branching ratio is already dominant at $n=1$.

The $Z'$ case, based on $Z'B(H^\dagger H)^n$, is more striking. To compare the minimal ($n=0$) and busy operators on the same footing, we fix the kinetic-mixing parameter as $\epsilon \equiv c_n v^2/(2\Lambda^2)$ and take $\Lambda=m_{Z'}$. Hence, the $q\bar{q}$ annihilation production rate of $Z'$ is identical in the two cases shown in \cref{fig:busy_fermion_projection_op}. 
The momentum-enhanced coupling strongly boosts the Higgs channel as discussed around \cref{eq:Z_decay_enhancement}. As a result, the $Z/\gamma+h$ final state becomes the leading discovery mode and yields a substantially stronger mass reach.  
This again illustrates the central point of this work: once higher-order Higgs structures are present, Higgs-rich final states can outperform the conventional non-Higgs channels.

\section{Summary and Outlook}
\label{sec:conclusion}
In this work, we identified a class of models in which Higgs final states can play a qualitatively different role from the conventional expectation based on $SU(2)$ symmetry and the Goldstone Equivalence Theorem. For heavy resonances with minimal couplings to the Higgs sector, Higgs and gauge-boson final states are typically correlated, leading to comparable sensitivities in channels such as $hh$, $ZZ$, and $WW$. We showed that this expectation can be violated once the resonance couples through higher-order Higgs structures. The key effect is that EWSB selectively amplifies Higgs-rich final states. In the scalar case, this can make $S\to hh$ one of the leading decay modes, while for $n\geq3$ multi-Higgs final states from heavy resonances can become important as well.

We further showed that the same underlying mechanism extends beyond the scalar case. For heavy fermionic resonances, Higgs-infused couplings can enhance the $ht$ channel relative to the corresponding gauge-boson modes. For vector resonances, analogous structures can make $Zh$ and $\gamma h$ especially prominent. Similar extensions also exist for spin-2 resonances, where higher-order Higgs operators naturally favor multi-Higgs final states. We also expect a similar lesson to hold for higher-spin resonances.  Taken together, these examples indicate that Higgs-rich channels are not merely complementary to conventional diboson searches but can instead provide the primary discovery mode for new resonances.

These observations have direct implications for collider searches and future theory developments. In the scalar case, our estimates show that di-Higgs searches can probe the busy operators more effectively than diboson channels across an important part of the parameter space. Moreover, the same framework opens a broader search direction in high-multiplicity Higgs final states, providing explicit motivation for resonant tri-Higgs and four-Higgs searches. For fermionic and vector resonances, the mechanism likewise motivates dedicated searches in $ht$, $Zh$, and $\gamma h$ final states. Therefore, our results point to Higgs final states as well-motivated and potentially leading discovery channels for new resonances.

\section{Acknowledgments}

We thank Keisuke Harigaya and Carlos Wagner for useful discussions. P.L. and Z.L. are supported by the Department of Energy under Grant No.~DE-SC0011842 at the University of Minnesota. P.L. is partly supported by a Doctoral Dissertation Fellowship at the University of Minnesota. Z.L. is supported in part by a Sloan Research Fellowship from the Alfred P. Sloan Foundation at the University of Minnesota. L.T.W. is supported by the Department of Energy under Grant No.~DE-SC0013642.


\bibliography{references}

@article{DiMicco:2019ngk,
    author = "Alison, J. and others",
    editor = "Di Micco, Biagio and Gouzevitch, Maxime and Mazzitelli, Javier and Vernieri, Caterina",
    title = "{Higgs boson potential at colliders: Status and perspectives}",
    eprint = "1910.00012",
    archivePrefix = "arXiv",
    primaryClass = "hep-ph",
    reportNumber = "FERMILAB-CONF-19-468-E-T, LHCXSWG-2019-005",
    doi = "10.1016/j.revip.2020.100045",
    journal = "Rev. Phys.",
    volume = "5",
    pages = "100045",
    year = "2020"
}

@article{An:2018dwb,
    author = "An, Fenfen and others",
    title = "{Precision Higgs physics at the CEPC}",
    eprint = "1810.09037",
    archivePrefix = "arXiv",
    primaryClass = "hep-ex",
    reportNumber = "FERMILAB-PUB-18-573-T",
    doi = "10.1088/1674-1137/43/4/043002",
    journal = "Chin. Phys. C",
    volume = "43",
    number = "4",
    pages = "043002",
    year = "2019"
}

@article{DiVita:2017vrr,
    author = "Di Vita, Stefano and Durieux, Gauthier and Grojean, Christophe and Gu, Jiayin and Liu, Zhen and Panico, Giuliano and Riembau, Marc and Vantalon, Thibaud",
    title = "{A global view on the Higgs self-coupling at lepton colliders}",
    eprint = "1711.03978",
    archivePrefix = "arXiv",
    primaryClass = "hep-ph",
    reportNumber = "DESY-17-131, FERMILAB-PUB-17-462-T",
    doi = "10.1007/JHEP02(2018)178",
    journal = "JHEP",
    volume = "02",
    pages = "178",
    year = "2018"
}

@article{Cohen:2025kwp,
    author = "Cohen, Timothy and McCullough, Matthew and Weiner, Neal",
    title = "{Powerful Yukawas}",
    eprint = "2512.04270",
    archivePrefix = "arXiv",
    primaryClass = "hep-ph",
    reportNumber = "CERN-TH-2025-227",
    month = "12",
    year = "2025"
}

@article{ATLAS:2023vdy,
    author = "Aad, Georges and others",
    collaboration = "ATLAS",
    title = "{Combination of Searches for Resonant Higgs Boson Pair Production Using pp Collisions at s=13{\,}{\,}TeV with the ATLAS Detector}",
    eprint = "2311.15956",
    archivePrefix = "arXiv",
    primaryClass = "hep-ex",
    reportNumber = "CERN-EP-2023-271",
    doi = "10.1103/PhysRevLett.132.231801",
    journal = "Phys. Rev. Lett.",
    volume = "132",
    number = "23",
    pages = "231801",
    year = "2024"
}

@article{ATLAS:2020fry,
    author = "Aad, Georges and others",
    collaboration = "ATLAS",
    title = "{Search for heavy diboson resonances in semileptonic final states in pp collisions at $\sqrt{s}=13$ TeV with the ATLAS detector}",
    eprint = "2004.14636",
    archivePrefix = "arXiv",
    primaryClass = "hep-ex",
    reportNumber = "CERN-EP-2020-049",
    doi = "10.1140/epjc/s10052-020-08554-y",
    journal = "Eur. Phys. J. C",
    volume = "80",
    number = "12",
    pages = "1165",
    year = "2020"
}

@article{Giudice:2008uua,
    author = "Giudice, Gian F. and Lebedev, Oleg",
    title = "{Higgs-dependent Yukawa couplings}",
    eprint = "0804.1753",
    archivePrefix = "arXiv",
    primaryClass = "hep-ph",
    reportNumber = "CERN-PH-TH-2008-069",
    doi = "10.1016/j.physletb.2008.05.062",
    journal = "Phys. Lett. B",
    volume = "665",
    pages = "79--85",
    year = "2008"
}

@techreport{CMS-PAS-HIG-24-012,
  author      = "{CMS Collaboration}",
  title       = "{Search for nonresonant triple Higgs boson production in the six b-quark final state in proton-proton collisions at 13 TeV}",
  institution = "CERN",
  number      = "CMS-PAS-HIG-24-012",
  year        = "2025",
  url         = "https://cds.cern.ch/record/2945361"
}

@article{ATLAS:2024xcs,
    author = "Aad, Georges and others",
    collaboration = "ATLAS",
    title = "{Search for triple Higgs boson production in the 6b final state using pp collisions at s=13{\,}{\,}TeV with the ATLAS detector}",
    eprint = "2411.02040",
    archivePrefix = "arXiv",
    primaryClass = "hep-ex",
    reportNumber = "CERN-EP-2024-285",
    doi = "10.1103/PhysRevD.111.032006",
    journal = "Phys. Rev. D",
    volume = "111",
    number = "3",
    pages = "032006",
    year = "2025"
}

@article{ATLAS:2022enb,
    author = "Aad, Georges and others",
    collaboration = "ATLAS",
    title = "{Search for heavy resonances decaying into a $Z$ or $W$ boson and a Higgs boson in final states with leptons and $b$-jets in $139~$fb$^{-1}$ of $pp$ collisions at $\sqrt{s}=13~$TeV with the ATLAS detector}",
    eprint = "2207.00230",
    archivePrefix = "arXiv",
    primaryClass = "hep-ex",
    reportNumber = "CERN-EP-2022-115",
    doi = "10.1007/JHEP06(2023)016",
    journal = "JHEP",
    volume = "06",
    pages = "016",
    year = "2023"
}

@article{ATLAS:2020jeb,
    author = "Aad, Georges and others",
    collaboration = "ATLAS",
    title = "{Search for heavy resonances decaying into a photon and a hadronically decaying Higgs boson in $pp$ collisions at $\sqrt{s}=13$ TeV with the ATLAS detector}",
    eprint = "2008.05928",
    archivePrefix = "arXiv",
    primaryClass = "hep-ex",
    reportNumber = "CERN-EP-2020-128, CERN preprint ID: CERN-EP-2020-128",
    doi = "10.1103/PhysRevLett.125.251802",
    journal = "Phys. Rev. Lett.",
    volume = "125",
    pages = "251802",
    year = "2020"
}

@article{CMS:2022fck,
    author = "Tumasyan, Armen and others",
    collaboration = "CMS",
    title = "{Search for pair production of vector-like quarks in leptonic final states in proton-proton collisions at $ \sqrt{s} $ = 13 TeV}",
    eprint = "2209.07327",
    archivePrefix = "arXiv",
    primaryClass = "hep-ex",
    reportNumber = "CMS-B2G-20-011, CERN-EP-2022-175",
    doi = "10.1007/JHEP07(2023)020",
    journal = "JHEP",
    volume = "07",
    pages = "020",
    year = "2023"
}

@article{Cepeda:2019klc,
    author = "Cepeda, M. and others",
    editor = "Dainese, Andrea and Mangano, Michelangelo and Meyer, Andreas B. and Nisati, Aleandro and Salam, Gavin and Vesterinen, Mika Anton",
    title = "{Report from Working Group 2}: {Higgs Physics at the HL-LHC and HE-LHC}",
    eprint = "1902.00134",
    archivePrefix = "arXiv",
    primaryClass = "hep-ph",
    reportNumber = "CERN-LPCC-2018-04",
    doi = "10.23731/CYRM-2019-007.221",
    journal = "CERN Yellow Rep. Monogr.",
    volume = "7",
    pages = "221--584",
    year = "2019"
}

@article{Maltoni:2024dpn,
    author = "Maltoni, Fabio and Ventura, Giuseppe and Vryonidou, Eleni",
    title = "{Impact of SMEFT renormalisation group running on Higgs production at the LHC}",
    eprint = "2406.06670",
    archivePrefix = "arXiv",
    primaryClass = "hep-ph",
    doi = "10.1007/JHEP12(2024)183",
    journal = "JHEP",
    volume = "12",
    pages = "183",
    year = "2024"
}

@article{CMS:2024phk,
    author = "Hayrapetyan, Aram and others",
    collaboration = "CMS",
    title = "{Searches for Higgs boson production through decays of heavy resonances}",
    eprint = "2403.16926",
    archivePrefix = "arXiv",
    primaryClass = "hep-ex",
    reportNumber = "CMS-B2G-23-002, CERN-EP-2024-062",
    doi = "10.1016/j.physrep.2024.09.004",
    journal = "Phys. Rept.",
    volume = "1115",
    pages = "368--447",
    year = "2025"
}

@article{CMS:2019kaf,
    author = "Sirunyan, Albert M and others",
    collaboration = "CMS",
    title = "{Combination of CMS searches for heavy resonances decaying to pairs of bosons or leptons}",
    eprint = "1906.00057",
    archivePrefix = "arXiv",
    primaryClass = "hep-ex",
    reportNumber = "CMS-B2G-18-006, CERN-EP-2019-110",
    doi = "10.1016/j.physletb.2019.134952",
    journal = "Phys. Lett. B",
    volume = "798",
    pages = "134952",
    year = "2019"
}

@article{Xiao:2014kba,
    author = "Xiao, Ming-Lei and Yu, Jiang-Hao",
    title = "{Stabilizing electroweak vacuum in a vectorlike fermion model}",
    eprint = "1404.0681",
    archivePrefix = "arXiv",
    primaryClass = "hep-ph",
    reportNumber = "UTTG-10-14, TCC-011-14",
    doi = "10.1103/PhysRevD.90.014007",
    journal = "Phys. Rev. D",
    volume = "90",
    number = "1",
    pages = "014007",
    year = "2014",
    note = "[Addendum: Phys.Rev.D 90, 019901 (2014)]"
}

@article{Aguilar-Saavedra:2013qpa,
    author = "Aguilar-Saavedra, J. A. and Benbrik, R. and Heinemeyer, S. and P{\'e}rez-Victoria, M.",
    title = "{Handbook of vectorlike quarks: Mixing and single production}",
    eprint = "1306.0572",
    archivePrefix = "arXiv",
    primaryClass = "hep-ph",
    doi = "10.1103/PhysRevD.88.094010",
    journal = "Phys. Rev. D",
    volume = "88",
    number = "9",
    pages = "094010",
    year = "2013"
}

@article{DeSimone:2012fs,
    author = "De Simone, Andrea and Matsedonskyi, Oleksii and Rattazzi, Riccardo and Wulzer, Andrea",
    title = "{A First Top Partner Hunter's Guide}",
    eprint = "1211.5663",
    archivePrefix = "arXiv",
    primaryClass = "hep-ph",
    reportNumber = "CERN-PH-TH-2012-323, SISSA-31-2012-EP",
    doi = "10.1007/JHEP04(2013)004",
    journal = "JHEP",
    volume = "04",
    pages = "004",
    year = "2013"
}

@article{Pappadopulo:2014qza,
    author = "Pappadopulo, Duccio and Thamm, Andrea and Torre, Riccardo and Wulzer, Andrea",
    title = "{Heavy Vector Triplets: Bridging Theory and Data}",
    eprint = "1402.4431",
    archivePrefix = "arXiv",
    primaryClass = "hep-ph",
    doi = "10.1007/JHEP09(2014)060",
    journal = "JHEP",
    volume = "09",
    pages = "060",
    year = "2014"
}

@article{Antipin:2007pi,
    author = "Antipin, Oleg and Atwood, David and Soni, Amarjit",
    title = "{Search for RS gravitons via W(L)W(L) decays}",
    eprint = "0711.3175",
    archivePrefix = "arXiv",
    primaryClass = "hep-ph",
    reportNumber = "BNL-HET-07-20",
    doi = "10.1016/j.physletb.2008.07.009",
    journal = "Phys. Lett. B",
    volume = "666",
    pages = "155--161",
    year = "2008"
}

@article{Agashe:2007zd,
    author = "Agashe, Kaustubh and Davoudiasl, Hooman and Perez, Gilad and Soni, Amarjit",
    title = "{Warped Gravitons at the LHC and Beyond}",
    eprint = "hep-ph/0701186",
    archivePrefix = "arXiv",
    reportNumber = "SU-4252-843, BNL-HET-07-3, YITP-SB-07-02",
    doi = "10.1103/PhysRevD.76.036006",
    journal = "Phys. Rev. D",
    volume = "76",
    pages = "036006",
    year = "2007"
}

@article{ATLAS:2019nat,
    author = "Aad, Georges and others",
    collaboration = "ATLAS",
    title = "{Search for diboson resonances in hadronic final states in 139 fb$^{-1}$ of $pp$ collisions at $\sqrt{s} = 13$ TeV with the ATLAS detector}",
    eprint = "1906.08589",
    archivePrefix = "arXiv",
    primaryClass = "hep-ex",
    reportNumber = "CERN-EP-2019-044",
    doi = "10.1007/JHEP09(2019)091",
    journal = "JHEP",
    volume = "09",
    pages = "091",
    year = "2019",
    note = "[Erratum: JHEP 06, 042 (2020)]"
}

@article{ATLAS:2017tlw,
    author = "Aaboud, M. and others",
    collaboration = "ATLAS",
    title = "{Search for heavy ZZ resonances in the $\ell ^+\ell ^-\ell ^+\ell ^-$ and $\ell ^+\ell ^-\nu \bar{\nu }$ final states using proton{\textendash}proton collisions at $\sqrt{s}= 13$   $\text {TeV}$ with the ATLAS detector}",
    eprint = "1712.06386",
    archivePrefix = "arXiv",
    primaryClass = "hep-ex",
    reportNumber = "CERN-EP-2017-251",
    doi = "10.1140/epjc/s10052-018-5686-3",
    journal = "Eur. Phys. J. C",
    volume = "78",
    number = "4",
    pages = "293",
    year = "2018"
}

@article{CMS:2022pjv,
    author = "Tumasyan, Armen and others",
    collaboration = "CMS",
    title = "{Search for new heavy resonances decaying to WW, WZ, ZZ, WH, or ZH boson pairs in the all-jets final state in proton-proton collisions at s=13TeV}",
    eprint = "2210.00043",
    archivePrefix = "arXiv",
    primaryClass = "hep-ex",
    reportNumber = "CMS-B2G-20-009, CERN-EP-2022-152",
    doi = "10.1016/j.physletb.2023.137813",
    journal = "Phys. Lett. B",
    volume = "844",
    pages = "137813",
    year = "2023"
}

@article{CMS:2018dff,
    author = "Sirunyan, Albert M and others",
    collaboration = "CMS",
    title = "{Search for a heavy resonance decaying to a pair of vector bosons in the lepton plus merged jet final state at $ \sqrt{s}=13 $ TeV}",
    eprint = "1802.09407",
    archivePrefix = "arXiv",
    primaryClass = "hep-ex",
    reportNumber = "CMS-B2G-16-029, CERN-EP-2018-015",
    doi = "10.1007/JHEP05(2018)088",
    journal = "JHEP",
    volume = "05",
    pages = "088",
    year = "2018"
}

@article{Babu:1999me,
    author = "Babu, K. S. and Nandi, S.",
    title = "{Natural fermion mass hierarchy and new signals for the Higgs boson}",
    eprint = "hep-ph/9907213",
    archivePrefix = "arXiv",
    reportNumber = "OSU-HEP-99-07",
    doi = "10.1103/PhysRevD.62.033002",
    journal = "Phys. Rev. D",
    volume = "62",
    pages = "033002",
    year = "2000"
}

@article{ATLAS:2018ziw,
    author = "Aaboud, Morad and others",
    collaboration = "ATLAS",
    title = "{Combination of the searches for pair-produced vector-like partners of the third-generation quarks at $\sqrt{s} =$ 13 TeV with the ATLAS detector}",
    eprint = "1808.02343",
    archivePrefix = "arXiv",
    primaryClass = "hep-ex",
    reportNumber = "CERN-EP-2018-205",
    doi = "10.1103/PhysRevLett.121.211801",
    journal = "Phys. Rev. Lett.",
    volume = "121",
    number = "21",
    pages = "211801",
    year = "2018"
}

@article{CMS:2024hbn,
    author = "Hayrapetyan, Aram and others",
    collaboration = "CMS",
    title = "{Search for a heavy resonance decaying into a Z and a Higgs boson in events with an energetic jet and two electrons, two muons, or missing transverse momentum in proton-proton collisions at $ \sqrt{s} $ = 13 TeV}",
    eprint = "2411.00202",
    archivePrefix = "arXiv",
    primaryClass = "hep-ex",
    reportNumber = "CMS-B2G-23-008, CERN-EP-2024-214",
    doi = "10.1007/JHEP02(2025)089",
    journal = "JHEP",
    volume = "02",
    pages = "089",
    year = "2025"
}

@article{CMS:2025rqr,
    author = "Hayrapetyan, Aram and others",
    collaboration = "CMS",
    title = "{Search for heavy H$γ$ and Z$γ$ resonances with a bottom quark-antiquark pair in the final state in proton-proton collisions at $\sqrt{s}$ = 13 TeV}",
    eprint = "2511.14583",
    archivePrefix = "arXiv",
    primaryClass = "hep-ex",
    reportNumber = "CMS-B2G-24-007, CERN-EP-2025-217",
    month = "11",
    year = "2025"
}

@article{CMS:2026nuu,
    author = "Aad, Georges and others",
    collaboration = "CMS, ATLAS",
    title = "{Combination of ATLAS and CMS searches for Higgs boson pair production at $\sqrt{s} = 13$ TeV}",
    eprint = "2602.23991",
    archivePrefix = "arXiv",
    primaryClass = "hep-ex",
    reportNumber = "CERN-EP-2026-011",
    month = "2",
    year = "2026"
}

@article{Buchalla:2013rka,
    author = "Buchalla, Gerhard and Cat{\`a}, Oscar and Krause, Claudius",
    title = "{Complete Electroweak Chiral Lagrangian with a Light Higgs at NLO}",
    eprint = "1307.5017",
    archivePrefix = "arXiv",
    primaryClass = "hep-ph",
    reportNumber = "LMU-ASC-42-13, LMU-ASC{\textasciitilde}42-13",
    doi = "10.1016/j.nuclphysb.2014.01.018",
    journal = "Nucl. Phys. B",
    volume = "880",
    pages = "552--573",
    year = "2014",
    note = "[Erratum: Nucl.Phys.B 913, 475--478 (2016)]"
}

@article{Kumar:1969jjy,
    author = "Kumar, R.",
    title = "{Covariant phase-space calculations of n-body decay and production processes}",
    doi = "10.1103/PhysRev.185.1865",
    journal = "Phys. Rev.",
    volume = "185",
    pages = "1865--1875",
    year = "1969"
}

@article{ATLAS:2012yve,
    author = "Aad, Georges and others",
    collaboration = "ATLAS",
    title = "{Observation of a new particle in the search for the Standard Model Higgs boson with the ATLAS detector at the LHC}",
    eprint = "1207.7214",
    archivePrefix = "arXiv",
    primaryClass = "hep-ex",
    reportNumber = "CERN-PH-EP-2012-218",
    doi = "10.1016/j.physletb.2012.08.020",
    journal = "Phys. Lett. B",
    volume = "716",
    pages = "1--29",
    year = "2012"
}

@article{CMS:2012qbp,
    author = "Chatrchyan, Serguei and others",
    collaboration = "CMS",
    title = "{Observation of a New Boson at a Mass of 125 GeV with the CMS Experiment at the LHC}",
    eprint = "1207.7235",
    archivePrefix = "arXiv",
    primaryClass = "hep-ex",
    reportNumber = "CMS-HIG-12-028, CERN-PH-EP-2012-220",
    doi = "10.1016/j.physletb.2012.08.021",
    journal = "Phys. Lett. B",
    volume = "716",
    pages = "30--61",
    year = "2012"
}

@article{CMS:2022dwd,
    author = "Tumasyan, Armen and others",
    collaboration = "CMS",
    title = "{A portrait of the Higgs boson by the CMS experiment ten years after the discovery.}",
    eprint = "2207.00043",
    archivePrefix = "arXiv",
    primaryClass = "hep-ex",
    reportNumber = "CMS-HIG-22-001, CERN-EP-2022-039",
    doi = "10.1038/s41586-022-04892-x",
    journal = "Nature",
    volume = "607",
    number = "7917",
    pages = "60--68",
    year = "2022",
    note = "[Erratum: Nature 623, (2023)]"
}

@article{ATLAS:2022vkf,
    author = "Aad, Georges and others",
    collaboration = "ATLAS",
    title = "{A detailed map of Higgs boson interactions by the ATLAS experiment ten years after the discovery}",
    eprint = "2207.00092",
    archivePrefix = "arXiv",
    primaryClass = "hep-ex",
    reportNumber = "CERN-EP-2022-057",
    doi = "10.1038/s41586-022-04893-w",
    journal = "Nature",
    volume = "607",
    number = "7917",
    pages = "52--59",
    year = "2022",
    note = "[Erratum: Nature 612, E24 (2022)]"
}

@article{Ruan:2021gap,
    author = "Ruan, Manqi and Fang, Yaquan and Li, Gang and Yu, Dan",
    title = "{CEPC Research Report: Higgs Physics Analysis}",
    eprint = "2107.09820",
    archivePrefix = "arXiv",
    primaryClass = "hep-ex",
    month = "7",
    year = "2021"
}

@article{Tian:2013yda,
    author = "Tian, Junping and Fujii, Keisuke",
    collaboration = "ILD",
    title = "{Measurement of Higgs couplings and self-coupling at the ILC}",
    eprint = "1311.6528",
    archivePrefix = "arXiv",
    primaryClass = "hep-ph",
    doi = "10.22323/1.180.0316",
    journal = "PoS",
    volume = "EPS-HEP2013",
    pages = "316",
    year = "2013"
}

@article{Barklow:2017suo,
    author = "Barklow, Tim and Fujii, Keisuke and Jung, Sunghoon and Karl, Robert and List, Jenny and Ogawa, Tomohisa and Peskin, Michael E. and Tian, Junping",
    title = "{Improved Formalism for Precision Higgs Coupling Fits}",
    eprint = "1708.08912",
    archivePrefix = "arXiv",
    primaryClass = "hep-ph",
    reportNumber = "DESY-17-120, KEK-PREPRINT-2017-22, SLAC-PUB-17129",
    doi = "10.1103/PhysRevD.97.053003",
    journal = "Phys. Rev. D",
    volume = "97",
    number = "5",
    pages = "053003",
    year = "2018"
}

@article{dEnterria:2017dac,
    author = "d'Enterria, David",
    title = "{Higgs physics at the Future Circular Collider}",
    eprint = "1701.02663",
    archivePrefix = "arXiv",
    primaryClass = "hep-ex",
    doi = "10.22323/1.282.0434",
    journal = "PoS",
    volume = "ICHEP2016",
    pages = "434",
    year = "2017"
}

@article{Maura:2025rcv,
    author = "Maura, Victor and Stefanek, Ben A. and You, Tevong",
    title = "{The Higgs Self-Coupling at FCC-ee}",
    eprint = "2503.13719",
    archivePrefix = "arXiv",
    primaryClass = "hep-ph",
    reportNumber = "KCL-PH-TH/2025-05",
    doi = "10.1103/wjcy-1qk6",
    journal = "Phys. Rev. Lett.",
    volume = "135",
    number = "14",
    pages = "141802",
    year = "2025"
}

@article{Forslund:2022xjq,
    author = "Forslund, Matthew and Meade, Patrick",
    title = "{High precision higgs from high energy muon colliders}",
    eprint = "2203.09425",
    archivePrefix = "arXiv",
    primaryClass = "hep-ph",
    reportNumber = "YITP-SB-22-11",
    doi = "10.1007/JHEP08(2022)185",
    journal = "JHEP",
    volume = "08",
    pages = "185",
    year = "2022"
}

@article{Forslund:2023reu,
    author = "Forslund, Matthew and Meade, Patrick",
    title = "{Precision Higgs width and couplings with a high energy muon collider}",
    eprint = "2308.02633",
    archivePrefix = "arXiv",
    primaryClass = "hep-ph",
    doi = "10.1007/JHEP01(2024)182",
    journal = "JHEP",
    volume = "01",
    pages = "182",
    year = "2024"
}

@article{Li:2024joa,
    author = "Li, Peiran and Liu, Zhen and Lyu, Kun-Feng",
    title = "{Higgs boson width and couplings at high energy muon colliders with forward muon detection}",
    eprint = "2401.08756",
    archivePrefix = "arXiv",
    primaryClass = "hep-ph",
    doi = "10.1103/PhysRevD.109.073009",
    journal = "Phys. Rev. D",
    volume = "109",
    number = "7",
    pages = "073009",
    year = "2024"
}

@article{deBlas:2022aow,
    author = "de Blas, Jorge and Gu, Jiayin and Liu, Zhen",
    title = "{Higgs boson precision measurements at a 125~GeV muon collider}",
    eprint = "2203.04324",
    archivePrefix = "arXiv",
    primaryClass = "hep-ph",
    doi = "10.1103/PhysRevD.106.073007",
    journal = "Phys. Rev. D",
    volume = "106",
    number = "7",
    pages = "073007",
    year = "2022"
}

@article{Froggatt:1978nt,
    author = "Froggatt, C. D. and Nielsen, Holger Bech",
    title = "{Hierarchy of Quark Masses, Cabibbo Angles and CP Violation}",
    reportNumber = "CERN-TH-2519",
    doi = "10.1016/0550-3213(79)90316-X",
    journal = "Nucl. Phys. B",
    volume = "147",
    pages = "277--298",
    year = "1979"
}

@article{Dolan:2012ac,
    author = "Dolan, Matthew J. and Englert, Christoph and Spannowsky, Michael",
    title = "{New Physics in LHC Higgs boson pair production}",
    eprint = "1210.8166",
    archivePrefix = "arXiv",
    primaryClass = "hep-ph",
    reportNumber = "IPPP-12-80, DCPT-12-160",
    doi = "10.1103/PhysRevD.87.055002",
    journal = "Phys. Rev. D",
    volume = "87",
    number = "5",
    pages = "055002",
    year = "2013"
}

@article{No:2013wsa,
    author = "No, Jose M. and Ramsey-Musolf, Michael",
    title = "{Probing the Higgs Portal at the LHC Through Resonant di-Higgs Production}",
    eprint = "1310.6035",
    archivePrefix = "arXiv",
    primaryClass = "hep-ph",
    doi = "10.1103/PhysRevD.89.095031",
    journal = "Phys. Rev. D",
    volume = "89",
    number = "9",
    pages = "095031",
    year = "2014"
}

@article{Chen:2014ask,
    author = "Chen, Chien-Yi and Dawson, S. and Lewis, I. M.",
    title = "{Exploring resonant di-Higgs boson production in the Higgs singlet model}",
    eprint = "1410.5488",
    archivePrefix = "arXiv",
    primaryClass = "hep-ph",
    doi = "10.1103/PhysRevD.91.035015",
    journal = "Phys. Rev. D",
    volume = "91",
    number = "3",
    pages = "035015",
    year = "2015"
}

@article{Kotwal:2016tex,
    author = "Kotwal, Ashutosh V. and Ramsey-Musolf, Michael J. and No, Jose Miguel and Winslow, Peter",
    title = "{Singlet-catalyzed electroweak phase transitions in the 100 TeV frontier}",
    eprint = "1605.06123",
    archivePrefix = "arXiv",
    primaryClass = "hep-ph",
    reportNumber = "ACFI-T16-12, FERMILAB-PUB-16-670",
    doi = "10.1103/PhysRevD.94.035022",
    journal = "Phys. Rev. D",
    volume = "94",
    number = "3",
    pages = "035022",
    year = "2016"
}

@article{Lewis:2017dme,
    author = "Lewis, Ian M. and Sullivan, Matthew",
    title = "{Benchmarks for Double Higgs Production in the Singlet Extended Standard Model at the LHC}",
    eprint = "1701.08774",
    archivePrefix = "arXiv",
    primaryClass = "hep-ph",
    doi = "10.1103/PhysRevD.96.035037",
    journal = "Phys. Rev. D",
    volume = "96",
    number = "3",
    pages = "035037",
    year = "2017"
}

@article{Lykken:2008bw,
    author = "Lykken, J. D. and Murdock, Z. and Nandi, S.",
    title = "{A light scalar as the messenger of electroweak and flavor symmetry breaking}",
    eprint = "0812.1826",
    archivePrefix = "arXiv",
    primaryClass = "hep-ph",
    reportNumber = "FERMILAB-PUB-08-432-T, OSU-HEP-08-08",
    doi = "10.1103/PhysRevD.79.075014",
    journal = "Phys. Rev. D",
    volume = "79",
    pages = "075014",
    year = "2009"
}

@article{Gavela:2016vte,
    author = "Gavela, M. B. and Kanshin, K. and Machado, P. A. N. and Saa, S.",
    title = "{The linear{\textendash}non-linear frontier for the Goldstone Higgs}",
    eprint = "1610.08083",
    archivePrefix = "arXiv",
    primaryClass = "hep-ph",
    reportNumber = "DFPD-2016-TH-17, FERMILAB-PUB-16-471-T, FTUAM-16-39, IFT-UAM-CSIC-16-107",
    doi = "10.1140/epjc/s10052-016-4541-7",
    journal = "Eur. Phys. J. C",
    volume = "76",
    number = "12",
    pages = "690",
    year = "2016"
}

@article{Buchalla:2016bse,
    author = "Buchalla, G. and Cata, O. and Celis, A. and Krause, C.",
    title = "{Standard Model Extended by a Heavy Singlet: Linear vs. Nonlinear EFT}",
    eprint = "1608.03564",
    archivePrefix = "arXiv",
    primaryClass = "hep-ph",
    reportNumber = "LMU-ASC-35-16",
    doi = "10.1016/j.nuclphysb.2017.02.006",
    journal = "Nucl. Phys. B",
    volume = "917",
    pages = "209--233",
    year = "2017"
}

@article{Hally:2012pu,
    author = "Hally, Katy and Logan, Heather E. and Pilkington, Terry",
    title = "{Constraints on large scalar multiplets from perturbative unitarity}",
    eprint = "1202.5073",
    archivePrefix = "arXiv",
    primaryClass = "hep-ph",
    doi = "10.1103/PhysRevD.85.095017",
    journal = "Phys. Rev. D",
    volume = "85",
    pages = "095017",
    year = "2012"
}

@article{Logan:2015xpa,
    author = "Logan, Heather E. and Rentala, Vikram",
    title = "{All the generalized Georgi-Machacek models}",
    eprint = "1502.01275",
    archivePrefix = "arXiv",
    primaryClass = "hep-ph",
    doi = "10.1103/PhysRevD.92.075011",
    journal = "Phys. Rev. D",
    volume = "92",
    number = "7",
    pages = "075011",
    year = "2015"
}

@article{Hisano:2013sn,
    author = "Hisano, Junji and Tsumura, Koji",
    title = "{Higgs boson mixes with an SU(2) septet representation}",
    eprint = "1301.6455",
    archivePrefix = "arXiv",
    primaryClass = "hep-ph",
    reportNumber = "IPMU13-0022",
    doi = "10.1103/PhysRevD.87.053004",
    journal = "Phys. Rev. D",
    volume = "87",
    pages = "053004",
    year = "2013"
}

@article{Alvarado:2014jva,
    author = "Alvarado, C. and Lehman, L. and Ostdiek, B.",
    title = "{Surveying the Scope of the $SU(2)_L$ Scalar Septet Sector}",
    eprint = "1404.3208",
    archivePrefix = "arXiv",
    primaryClass = "hep-ph",
    doi = "10.1007/JHEP05(2014)150",
    journal = "JHEP",
    volume = "05",
    pages = "150",
    year = "2014"
}

@article{Dawson:2017vgm,
    author = "Dawson, Sally and Murphy, Christopher W.",
    title = "{Standard Model EFT and Extended Scalar Sectors}",
    eprint = "1704.07851",
    archivePrefix = "arXiv",
    primaryClass = "hep-ph",
    doi = "10.1103/PhysRevD.96.015041",
    journal = "Phys. Rev. D",
    volume = "96",
    number = "1",
    pages = "015041",
    year = "2017"
}

\clearpage
\appendix
\onecolumngrid

\section{Branching Ratio of the Busy Scalar}
\label{sec:br}
In this section, we present the tree-level branching ratios generated by the busy-scalar operator $S(H^\dagger H)^n$, including final states up to four-body decays. Higher-multiplicity modes can be considered, but for $m_S$ around the TeV scale, they are more phase-space suppressed.

\subsection{Calculation in Goldstone Basis}
\begin{figure}
    \centering
    \includegraphics[width=0.497\linewidth]{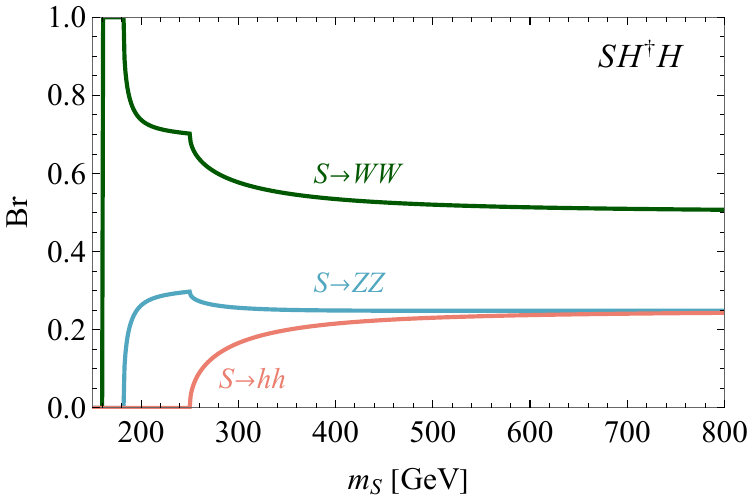}\\
    \vspace{5ex}
    \includegraphics[width=0.497\linewidth]{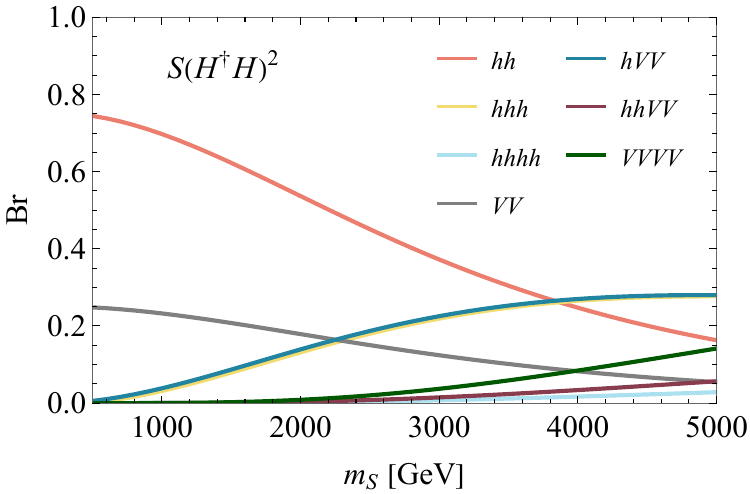}
    \includegraphics[width=0.497\linewidth]{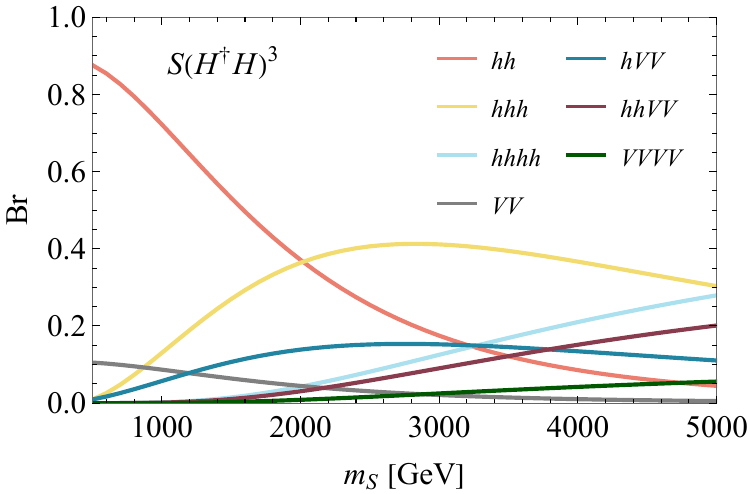}
    \caption{Branching ratios (tree level) for $SH^\dagger H$, $S(H^\dagger H)^2$ and $S(H^\dagger H)^3$, including final states with up to four particles. Here, $V$ denotes a $W$ or $Z$ boson. 
    }
    \label{fig:Br_tree_level}
\end{figure}

The relevant interactions follow from
\begin{align}
    \mathcal{L}\supset & \frac{c_n}{\Lambda^{2n-3}}S(H^\dagger H)^n \\
    =&\frac{c_n}{2^n\Lambda^{2n-3}}S\left[(v+h)^2+\sum_{i-0}^2 a_i^2\right]^n
    =\frac{c_n}{2^n\Lambda^{2n-3}}S\left[\sum_{j=0}^n\binom{n}{j} (\Sigma_i  a_i^2)^j(v+h)^{2(n-j)}  \right]\\
    \supset& \frac{c_n}{2^n\Lambda^{2n-3}}S\left[ \binom{2n}{2}h^2 v^{2n-2} + \binom{2n}{3}h^3 v^{2n-3} +\binom{2n}{4}h^4 v^{2n-4} + \cdots  \right . \nonumber \\
    &\qquad\qquad~~ +n(\Sigma_i  a_i^2)v^{2n-2}+2n(n-1)(\Sigma_i  a_i^2)hv^{2n-3}+n\binom{2n-2}{2}(\Sigma_i  a_i^2)h^2v^{2n-4}+\cdots \nonumber \\
    &~\qquad\qquad ~\left. +\binom{n}{2} (\Sigma_i  a_i^2)^2 v^{2(n-2)}+\cdots \right]~.
\end{align}
From these interactions, the amplitude ratios among different decay channels are
\begin{align}
    &\frac{\mathcal{M}(S\to hh)}{\mathcal{M}(S\to a_i a_i)}=(2n-1),\quad \frac{\mathcal{M}(S\to hhh)}{\mathcal{M}(S\to h a_i a_i)}=(2n-1), \nonumber \\
    &\mathcal{M}(S\to 4h): \mathcal{M}(S\to hh a_i a_i):\mathcal{M}(4a_i):\mathcal{M}(a_ia_i a_j a_j)=(2n-1)(2n-3):(2n-3):3:1~.
\end{align}
Hence, the ratios between partial decay rates are~\footnote{One needs to sum $a_1 a_1$ and $a_2 a_2$ partial widths to get the corresponding $W^+W^-$ partial width in the high energy limit.}
\begin{align}
    &\frac{\Gamma(S\to hh)}{\Gamma(S\to a_i a_i)}=(2n-1)^2,\quad \frac{\Gamma(S\to hhh)}{\Gamma(S\to h a_i a_i)}=(2n-1)^2/3, \nonumber \\
    &\Gamma(S\to 4h): \Gamma(S\to hh a_i a_i):\Gamma(4a_i):\Gamma(a_ia_i a_j a_j)=(2n-1)^2 (2n-3)^2/6:(2n-3)^2:3/2:1~.
\end{align}
Including the phase-space suppression from massive final states and identical particle suppression, 
we show the resulting branching fractions in \cref{fig:Br_tree_level}. Importantly, the branching ratios of multi-Higgs final states are enhanced with larger $n$, providing unique signatures at colliders.

\subsection{Cross-check in Unitary Gauge}
\label{app:unitary_gauge}

It is also useful to estimate the amplitudes directly in unitary gauge. In unitary gauge,
\begin{align}
    H=\frac{1}{\sqrt2}\begin{pmatrix}0\\ v+h\end{pmatrix},
    \qquad
    H^\dagger H=\frac{(v+h)^2}{2},
\end{align}
so that
\begin{align}
    \mathcal L \supset \frac{c_n}{\Lambda^{2n-3}}S(H^\dagger H)^n
    =\frac{c_n}{2^n\Lambda^{2n-3}}S(v+h)^{2n}.
\end{align}
Expanding the operator gives
\begin{align}
    \mathcal L \supset
    \frac{c_n}{2^n\Lambda^{2n-3}}
    \left[
        (2n)v^{2n-1}Sh
        +\binom{2n}{2}v^{2n-2}Sh^2
        +\binom{2n}{3}v^{2n-3}Sh^3
        +\cdots
    \right].
\end{align}

The term proportional to $S v^{2n}$ is a tadpole term. In principle, one can include a linear tadpole counterterm to make sure $\langle S\rangle=0$, so that $S$ is defined around the true vacuum. This does not affect the branching-ratio pattern discussed here.

The channels involving electroweak gauge bosons arise through the $Sh$ mixing term together with
the standard-model Higgs couplings
\begin{align}
    \mathcal L_{\rm SM}\supset
    \frac{2m_W^2}{v}hW_\mu^+W^{-\mu}
    +\frac{m_Z^2}{v}hZ_\mu Z^\mu
    +\frac{m_W^2}{v^2}h^2W_\mu^+W^{-\mu}
    +\frac{m_Z^2}{2v^2}h^2Z_\mu Z^\mu+\cdots .
\end{align}
In the heavy-scalar limit $m_S\gg m_h,m_W,m_Z$, the gauge-boson channels are dominated by
longitudinal modes. Using
\begin{align}
    \epsilon_L(p_1)\cdot\epsilon_L(p_2)\simeq \frac{p_1\cdot p_2}{m_V^2},
\end{align}
one finds for the two-body channels
\begin{align}
    \mathcal M(S\to W_L^+W_L^-)
    &\approx
    \frac{c_n}{2^n\Lambda^{2n-3}}
    (2n)v^{2n-1}
    \frac{1}{m_S^2-m_h^2}
    \left(\frac{2m_W^2}{v}\right)
    \left(\frac{m_S^2}{2m_W^2}\right)
    \nonumber\\
    &\approx
    \frac{c_n}{2^n\Lambda^{2n-3}}
    \,2n\,v^{2n-2},
    \\
    \mathcal M(S\to Z_LZ_L)
    &\approx
    \frac{c_n}{2^n\Lambda^{2n-3}}
    (2n)v^{2n-1}
    \frac{1}{m_S^2-m_h^2}
    \left(\frac{2m_Z^2}{v}\right)
    \left(\frac{m_S^2}{2m_Z^2}\right)
    \nonumber\\
    &\approx
    \frac{c_n}{2^n\Lambda^{2n-3}}
    \,2n\,v^{2n-2},
    \\
    \mathcal M(S\to hh)
    &\approx
    \frac{c_n}{2^n\Lambda^{2n-3}}
    \,(2n)(2n-1)\,v^{2n-2},
\end{align}
where the mixing-induced contribution to \(S\to hh\) through the SM Higgs cubic coupling is
suppressed by \(m_h^2/m_S^2\) and has been neglected.

For the three-body channels, the direct \(Sh^3\) interaction gives
\begin{align}
    \mathcal M(S\to hhh)
    \approx
    \frac{c_n}{2^n\Lambda^{2n-3}}
    \,(2n)(2n-1)(2n-2)\,v^{2n-3}.
\end{align}
For \(S\to hVV\), the leading longitudinal contributions come from the \(Sh^2\) interaction followed
by \(h^\ast\to VV\), together with the \(Sh\) mixing followed by the \(hhVV\) contact term. Their sum
scales as
\begin{align}
    \mathcal M(S\to hW_L^+W_L^-)
    &\approx
    \frac{c_n}{2^n\Lambda^{2n-3}}
    \,2n(2n-2)\,v^{2n-3},
    \\
    \mathcal M(S\to hZ_LZ_L)
    &\approx
    \frac{c_n}{2^n\Lambda^{2n-3}}
    \,2n(2n-2)\,v^{2n-3}.
\end{align}
Likewise, the four-body amplitudes in unitary gauge will also reproduce the same combinatorial pattern as in the Goldstone-basis analysis, once the longitudinal electroweak gauge bosons are identified with the corresponding Goldstone modes in the high-energy limit. We therefore do not present the explicit expressions here. One can also check that the operators $\partial_\mu S\,\partial^\mu(H^\dagger H)$, 
$S|D_\mu H|^2$, and $S H^\dagger H$ all lead to the same branching-ratio pattern
in the high-energy limit,
\begin{align}
    \mathrm{Br}(S\to hh):\mathrm{Br}(S\to W^+W^-):\mathrm{Br}(S\to ZZ)=1:2:1~.
\end{align}

\section{Loop Effect from EFT Operator}
\label{sec:loop_eft}
\begin{figure}
    \centering
    \includegraphics[width=0.497\linewidth]{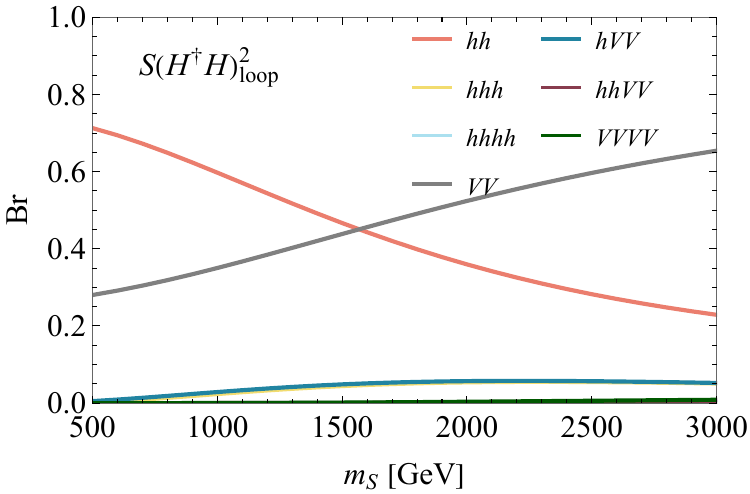}
    \includegraphics[width=0.497\linewidth]{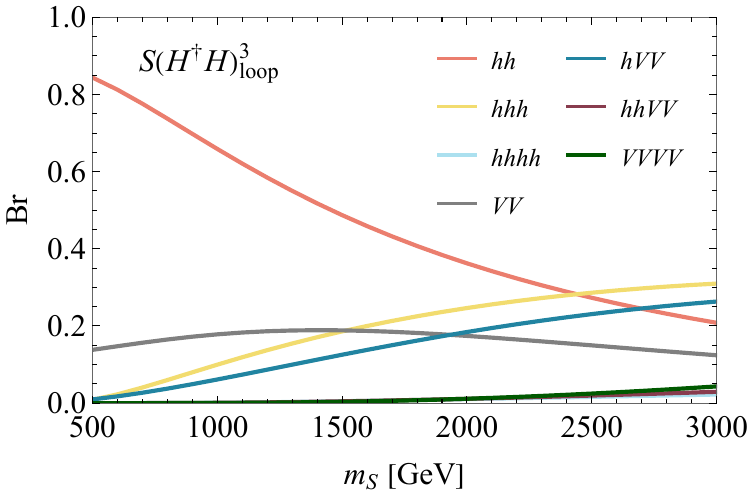}
    
    \caption{Branching ratios (1-loop level) for $S(H^\dagger H)^2$ and $S(H^\dagger H)^3$, including final states with up to four particles. Here, $V$ denotes a $W$ or $Z$ boson.
    }
    \label{fig:Br_loop}
\end{figure}

Within the class of ``busy'' operators, a lower-dimensional interaction is generically induced by contracting one pair of $H^\dagger H$ fields in the higher-dimensional operator $\frac{c_n}{\Lambda^{2n-3}} S (H^\dagger H)^n$. The corresponding loop integral is quadratically divergent and is naturally cut off at $\Lambda$,
\begin{align}
I=\frac{n^2\Lambda^2}{16\pi^2},
\end{align}
where the factor $n^2$ accounts for the number of ways to close one $H^\dagger H$ pair. The induced correction to the lower-dimensional operator scales as
\begin{align}
\Delta\mathcal{O}_{n-1}\sim \frac{n^2\Lambda^2}{16\pi^2}\frac{c_n}{\Lambda^{2n-3}} S (H^\dagger H)^{n-1}.
\end{align}
Parametrically, this modifies the decay amplitudes by $\sim \frac{n^2}{16\pi^2}\frac{\Lambda^2}{v^2}$. For $n=2$, taking $\mathcal{O}_{n-1}$ to be dominated by the loop-induced contribution, the amplitudes for $S\to hh/W_LW_L/Z_LZ_L$ become
\begin{align}
\mathcal{M}(S\to hh)
&=c_2\frac{v^2}{\Lambda}\left[3+\frac{4}{16\pi^2}\frac{\Lambda^2}{v^2}\right],\\
\mathcal{M}(S\to W_LW_L)
=\mathcal{M}(S\to Z_LZ_L)&= c_2\frac{v^2}{\Lambda}\left[1+\frac{4}{16\pi^2}\frac{\Lambda^2}{v^2}\right].\label{eq:loop_correction_eft}
\end{align}
This estimate is purely EFT-based, but it suggests that the correction is sensitive to the heavy degrees of freedom that generate the operator. As discussed in the next section, in many UV completions the loop-induced contribution is proportional to the heavy mass scale with logarithmic running. The one-loop correction to the branching ratio generated by $S(H^\dagger H)^2$ and $S(H^\dagger H)^3$ is shown in \cref{fig:Br_loop}.

\section{UV Models for Busy Resonance}
\label{sec:UV_model}
In this appendix, we present several representative UV completions for the busy-scalar scenario. In general, many BSM frameworks~\cite{Babu:1999me,Giudice:2008uua,Lykken:2008bw,Cohen:2025kwp,Gavela:2016vte,Buchalla:2016bse,Buchalla:2013rka} can generate similar behavior and realize a busy resonance.
\subsubsection{Heavy Triplet Higgs Portal}
A simple model for the busy operators can be realized by integrating out a heavy Higgs triplet $\Phi$. The relevant Lagrangian could be
\begin{align}
    \mathcal{L} \supset \mu S\Phi^\dagger \Phi
    + \kappa \Phi^*_{ab} H^a H^b
    + M_\Phi^2 \Phi^\dagger \Phi + \text{h.c.},
\end{align}
where $a,b\in \{1,2\}$ are $SU(2)$ index, $\Phi_{ab}$ is a symmetric $SU(2)$ tensor field. The components of $\Phi$ can be identified as
\begin{align}
    &\Phi_{11}=\Phi^{++},~
    \Phi_{22}=\Phi^{0},\
    \Phi_{12}=\Phi_{21}=\frac{1}{\sqrt{2}}\Phi^{+}.
\end{align}
After integrating out the heavy field $\Phi$, an effective operator
\begin{align}
\mathcal{O}_\text{EFT}=\frac{|\kappa|^2\mu}{M_\Phi^4}S(H^\dagger H)^2
\end{align}
is generated (first panel of \cref{fig:SHHHH_diagram}). More generally, operators with higher powers of $(H^\dagger H)^n$ can be obtained by introducing heavy scalars in higher $SU(2)_L$ representations. 

Loop effects can induce lower-dimensional operators. In particular, contracting one $H$ with one $H^\dagger$ inside $S(H^\dagger H)^2$ generates $SH^\dagger H$ at one loop:
\begin{align}
    \mathcal{O}_\text{loop}\sim \frac{4}{16\pi^2}\frac{|\kappa|^2\mu}{M_\Phi^2}SH^\dagger H
\end{align}
where the factor 4 counts the number of ways to close the Higgs loop. After EWSB, the amplitude of $S\to hh$ from the induced $S H^\dagger H$ operator relative to that from $S(H^\dagger H)^2$ scales as
$\sim \frac{4}{16\pi^2}\frac{M_\Phi^2}{v^2}$  (similar to \cref{eq:loop_correction_eft}),
so for $M_\Phi \geq 2\pi v$ the lower-dimensional operator can dominate.

Note that there are many different possibilities of high representations of $SU(2)$ scalar extensions of the SM~\cite{Hally:2012pu,Logan:2015xpa,Hisano:2013sn,Alvarado:2014jva,Dawson:2017vgm}, and they face various direct and indirect constraints. We choose a simple version here to show that one can readily generate the Busy Higgs operators. For fuller model-space explorations and parameter-region studies, we leave these for future work.

\subsubsection{Fermion-Loop-Induced Busy Scalars}
\label{sec:fermion_loop}
Another compelling UV completion involves fermionic loops that require multiple Higgs insertions, analogous to ``flavoron" models where the Higgs acts as a flavor mediator.
Consider a set of vector-like fermions $\psi_i$ with charges arranged such that the lowest-dimension coupling to $S$ requires a chain of Higgs insertions.
Diagrams requiring $n$ Higgs fields to close the loop generate $S(H^\dagger H)^n$ operators with characteristic loop suppression.

\begin{figure}
    \centering
    \includegraphics[width=0.23\linewidth]{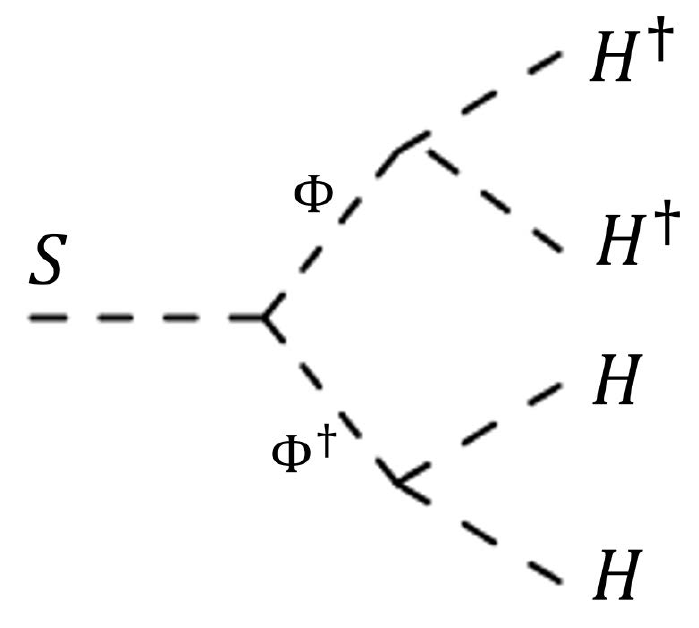}
    \includegraphics[width=0.3\linewidth]{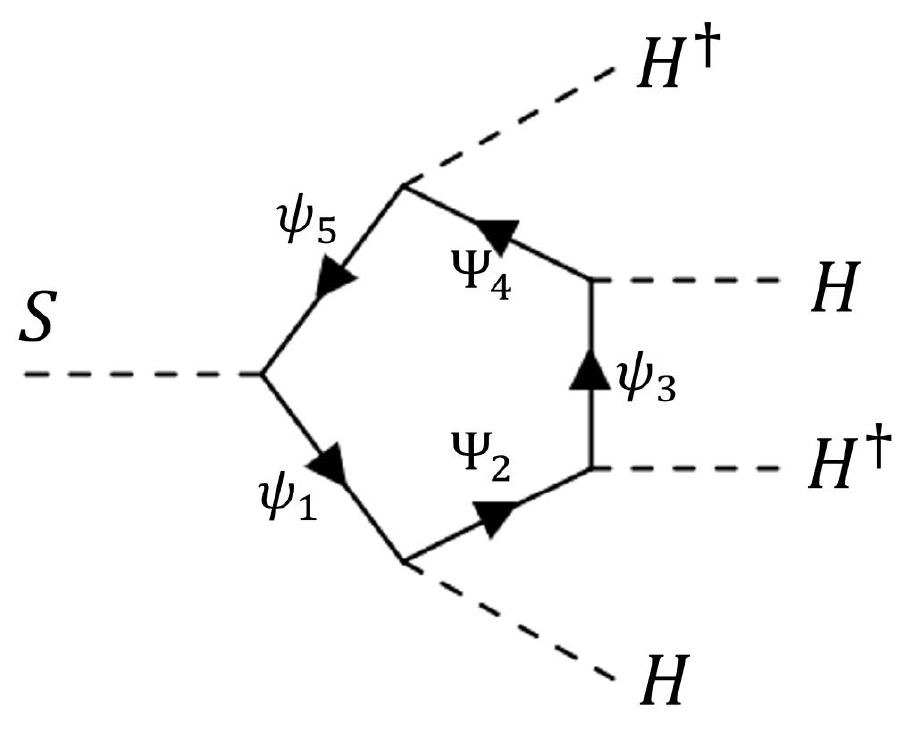}
    \includegraphics[width=0.3\linewidth]{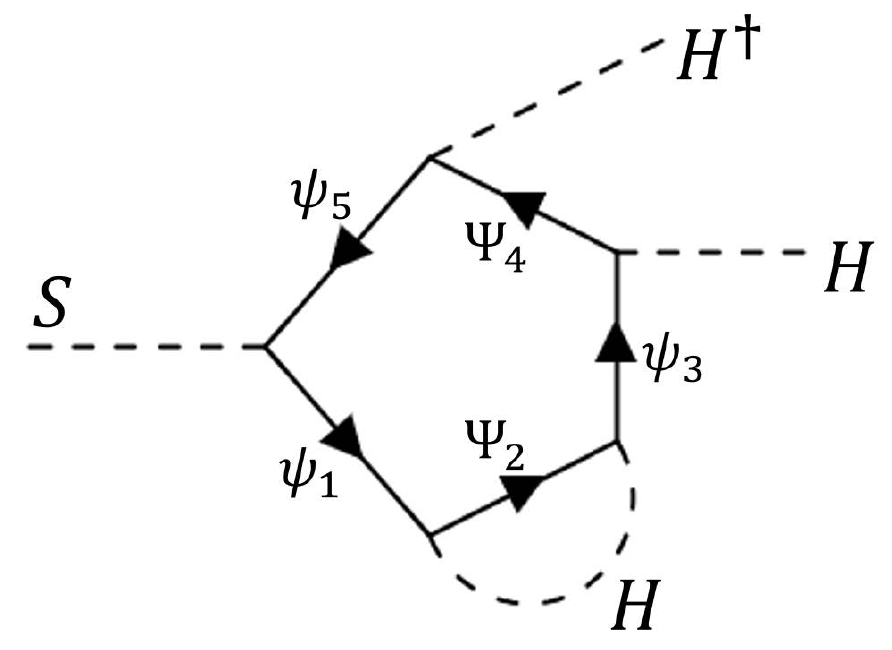}
    \caption{\textbf{First:} Effective operator $S (H^\dagger H)^2$ generated by heavy triplet scalar model. \textbf{Second:} Effective operator $S(H^\dagger H)^2$ generated by a fermion loop within loop-induced busy scalar model. \textbf{Third:} Effective operator $SH^\dagger H$ generated at two loops within fermion-loop-induced busy scalar model.}
    \label{fig:SHHHH_diagram}
\end{figure}

A concrete realization of the dimension-five operator $\frac{c_2}{\Lambda} S (H^\dagger H)^2$ can be inspired by the Giudice-Lebedev model~\cite{Giudice:2008uua}, introducing heavy fermions that mediate the interaction at one loop. As an explicit example, consider the following Lagrangian,
\begin{align}
    \mathcal{L}\supset&
    - (\overline{\Psi}_{2} H)\psi_{1}
    - \overline{\psi}_{3} (H^\dagger \Psi_{2})
    - (\overline{\Psi}_{4} H)\psi_{3} - \overline{\psi}_{5} (H^\dagger \Psi_{4})
    -S\overline{\psi}_{1}\psi_{5} 
    + \text{h.c.} \label{eq:fermion_loop}
\end{align}
where $\Psi_{2,4}$ are SM $SU(2)$ doublets heavy Dirac fermions, and $\psi_{1,3,5}$ are heavy SM-singlet Dirac fermions.
The operator $S(H^\dagger H)^2$ is then generated through the one-loop diagram shown in the second panel of \cref{fig:SHHHH_diagram}.

It is important to emphasize that, since $H^\dagger H$ is neutral under any global or gauge symmetry, the structure in \cref{eq:fermion_loop} is not protected by flavor symmetries such as the Froggatt–Nielsen mechanism~\cite{Froggatt:1978nt}. In contrast, in the MSSM or two-Higgs-doublet models (2HDMs), gauge-invariant combinations such as $H_u \epsilon H_d$ can carry non-zero charges and thus naturally play the role of a flavon~\cite{Giudice:2008uua}, allowing symmetry control over the effective operators.

As an explicit example, consider a 2HDM supplemented by heavy fermions,
\begin{align}
    \mathcal{L}\supset&
    - (\overline{\Psi}_{2} H_u)\psi_{1}
    - \overline{\psi}_{3} (\Psi_{2}^T \epsilon H_d)
    - (\overline{\Psi}_{4} H_u)\psi_{3} - \overline{\psi}_{5} (\Psi_{4}^T \epsilon H_d)
    -\Phi\overline{\psi}_{1}\psi_{5} 
    + \text{h.c.} \label{eq:fermion_loop_2HDM}
\end{align}
where $H_u,H_d, \Psi_{2,4}$ are $SU(2)$ doublets and $\Phi$ is a complex scalar field (SM singlet), $\psi_{1,3,5}$ are SM singlets.
All these heavy fermions are vector-like. Their $U(1)$ flavor charges are assigned as shown in \cref{tab:flavor_charge}. The $U(1)$ flavor symmetry ensures that the leading effective coupling between $\Phi$ and the Higgs fields takes the form $\Phi (H_u \epsilon H_d)^2$, which is generated via fermion loop.

\begin{table}[htbp]
\centering
\caption{Flavor charge assignment}
\label{tab:flavor_charge}
\begin{tabular}{lccccccccc}
\toprule
 & $\psi_1$ & $\Psi_2$ & $\psi_3$
& $\Psi_4$ & $\psi_5$ & $H_u$ & $H_d$ & $\Phi$ \\
$U(1)$ flavor charge & $1$ & $2$ & $3$ & $4$ & $5$ & $1$ & $1$ & $-4$\\
SM $U(1)_Y$ & $0$ & $\frac{1}{2}$ & $0$ & $\frac{1}{2}$ & $0$ & $\frac{1}{2}$ & $-\frac{1}{2}$ & $0$\\
SM $SU(2)_L$ & $0$ & $\frac{1}{2}$ & $0$ & $\frac{1}{2}$ & $0$ & $\frac{1}{2}$ & $\frac{1}{2}$ & $0$\\
\end{tabular}
\end{table}

We now discuss the SM-like Higgs components contained in $H_u \epsilon H_d$. The two Higgs doublets are parametrized by
\begin{align}
    H_u = \frac{1}{\sqrt{2}}
    \begin{pmatrix}
        v_u+ \text{Re}H_u^0 + i \text{Im}H_u^0 \\
        \sqrt{2}H_u^-
    \end{pmatrix},~    
    H_d = \frac{1}{\sqrt{2}}
    \begin{pmatrix}
        \sqrt{2}H_d^+\\
        v_d+ \text{Re}H_d^0 + i \text{Im}H_d^0 
    \end{pmatrix}.
\end{align}
In the mass-eigenstate basis, the CP-even neutral scalars $h$ and $H$, CP-odd neutral scalar $A$, charged scalars $H^\pm$ and the Goldstone modes $\pi^{0,\pm}$ can be parametrized by 
\begin{align}
    \begin{pmatrix}
        h \\
        H
    \end{pmatrix} = 
    \begin{pmatrix}
        c_\alpha & -s_\alpha \\
        s_\alpha & c_\alpha
    \end{pmatrix}
    \begin{pmatrix}
        \text{Re}H_u^0 \\
        \text{Re}H_d^0
    \end{pmatrix},~
    \begin{pmatrix}
        \pi^0 \\
        A
    \end{pmatrix} = 
    \begin{pmatrix}
        s_\beta & -c_\beta \\
        c_\beta & s_\beta
    \end{pmatrix}
    \begin{pmatrix}
        \text{Im}H_u^0 \\
        \text{Im}H_d^0
    \end{pmatrix},~
    \begin{pmatrix}
        \pi^- \\
        H^-
    \end{pmatrix} = 
    \begin{pmatrix}
        s_\beta & -c_\beta \\
        c_\beta & s_\beta
    \end{pmatrix}
    \begin{pmatrix}
        H_u^- \\
        {H_d^+}^*
    \end{pmatrix},
\end{align}
where $\tan\beta=v_u/v_d$.
For simplicity, we work in the decoupling limit, $\alpha=\beta-\pi/2$. In this limit, $H_u \epsilon H_d$ can be expanded as
\begin{align}
    H_u \epsilon H_d&=\frac{1}{4}\sin(2\beta) \left[(v+h)^2 + (\pi^0)^2 + 2\pi^-\pi^+\right] \nonumber \\
    &~~~- \frac{1}{2}v\cos(2\beta) H - \frac{1}{4}\sin(2\beta)[H^2+A^2] \nonumber \\
    &\quad + \frac{1}{2}\cos(2\beta)(\pi^0 A-hH)+\frac{i}{2}(vA+hA+\pi^0 H) \nonumber \\
    &\quad - \frac{1}{2}\sin(2\beta)H^-H^+ + \sin^2\beta~ (\pi^- H^+) - \cos^2\beta~ (\pi^+ H^-)~. \label{eq:HuHd}
\end{align}
Although the gauge-invariant combination $H_u \epsilon H_d$ contains multiple scalar components, the relevant low-energy dynamics is considerably simplified once the heavy Higgs states are kinematically inaccessible. In particular, assuming
\begin{align}
    M_H, M_A, M_{H^\pm} > M_\psi > m_\Phi~,
\end{align}
the heavy scalar $\Phi$ can decay only into the SM-like Higgs boson $h$ and the electroweak Goldstone modes $\pi^{0,\pm}$. As a result, among the various components appearing in the expansion of $H_u \epsilon H_d$, only the leading term in \cref{eq:HuHd} contributes in our scenario. Since the final EFT operator should be
\begin{align}
    \sim\frac{c_2}{(4\pi)^2}\frac{1}{M_\psi}\Phi(H^\dagger H)^2 + \text{h.c.}~,
\end{align}
the interaction effectively involves only the real component of $\Phi$ after integrating out the heavy modes and combining with the Hermitian conjugate, which is a consequence of decoupling/alignment limit. The resulting EFT structure is therefore equivalent to $S(H^\dagger H)^2$.\footnote{Once $H_u,H_d$ acquire vevs, the $U(1)_\text{flavor}$ symmetry is broken and the corresponding pseudo-Goldstone boson is $A$. One of the explicit broken terms in the potential is $m_{12}^2(H_u\epsilon H_d+\text{h.c.})$.}

Turning back to loop corrections, one can also generate $SH^\dagger H$ at two loops (third panel of \cref{fig:SHHHH_diagram}), involving five fermion propagators and one Higgs propagator. Since only an even number of gamma matrices contributes to the trace, the result is proportional to the heavy fermion mass $M_\psi$ and the loop integral is logarithmically divergent. Thus, the correction should scale as
\begin{align}
    \sim \frac{c_2}{(4\pi)^4}M_\psi \ln\left(\frac{\Lambda_\text{UV}^2}{m_S^2}\right) SH^\dagger H,
\end{align}
where $\Lambda_\text{UV}$ denotes the UV scale associated with the heavy 2HDM modes and the heavy fermions. This behavior is consistent with the EFT expectation in \cref{sec:loop_eft}.

\subsubsection{Scalar-Loop-Induced Busy Scalars}
\label{sec:scalar_loop}
The same idea can also be implemented with scalar loops. For example, consider the Lagrangian
\begin{align}
\mathcal{L}\supset \mu\Phi N_1^\dagger N_{5} + N_1 N_3^\dagger H_u\epsilon H_d + N_3 N_5^\dagger H_u\epsilon H_d + \text{h.c.}
\end{align}
where $\Phi$ and $N_i$ are complex scalars. Their $U(1)$ flavor charges are assigned as shown in \cref{tab:flavor_charge_scalar}. The heavy scalar $S$ is identified with the real component of $\Phi$. The induced operator is
\begin{align}
c_h\frac{\mu}{M_N^2}S(H^\dagger H)^2,
\end{align}
where $M_N$ denotes the masses of $N_i$.

\begin{table}[htbp]
\centering
\caption{Flavor charge assignment}
\label{tab:flavor_charge_scalar}
\begin{tabular}{lccccccc}
\toprule
 & $N_1$ & $N_3$ & $N_5$
& $H_u$ & $H_d$ & $\Phi$ \\
$U(1)$ flavor charge & $1$ & $3$ & $5$ & $1$ & $1$ & $-4$\\
SM $U(1)_Y$ & $0$ & $0$ & $0$ & $\frac{1}{2}$ & $-\frac{1}{2}$ & $0$ \\
SM $SU(2)_L$ & $0$ & $0$ & $0$ & $\frac{1}{2}$ & $\frac{1}{2}$ & $0$ \\
\end{tabular}
\end{table}

\begin{figure}[h]
    \centering
    \includegraphics[width=0.23\linewidth]{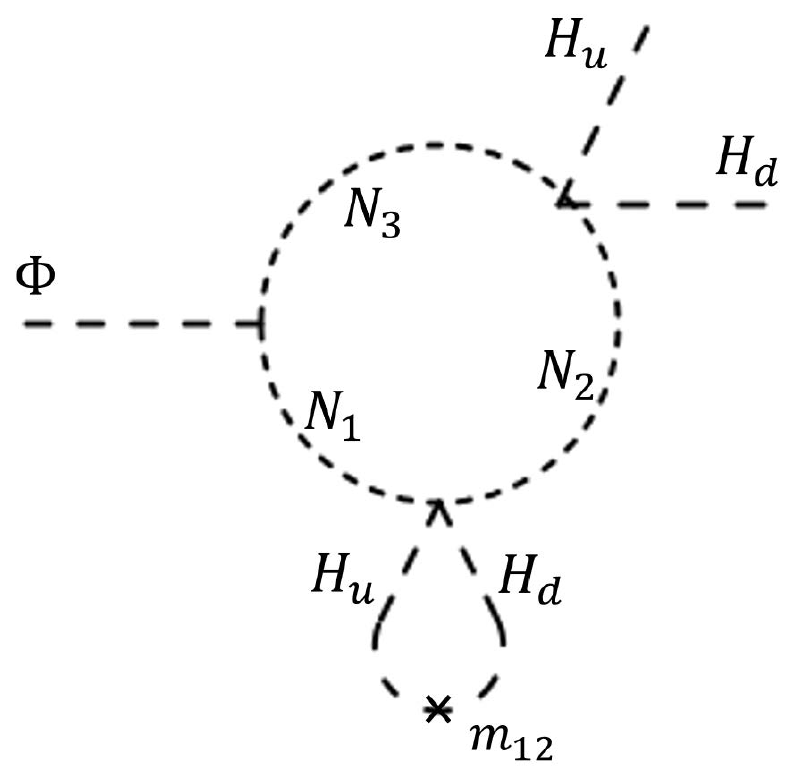}
    \caption{Loop correction with insertion of $m_{12}^2$ term to generate lower-dimensional operator within scalar-loop-induced busy scalar model.}
    \label{fig:SHH_loop_scalar}
\end{figure}

In this setup, the loop-induced lower-dimensional operator is directly sensitive to the $m_{12}^2$ term in a 2HDM (or the $B$ term in SUSY). This can be seen by closing the $H_u$ and $H_d$ lines through the insertion $m_{12}^2(H_u\epsilon H_d)$ shown in \cref{fig:SHH_loop_scalar}. The resulting correction scales as
\begin{align}
\sim \frac{1}{16\pi^2}m_{12}^2\ln \left(\frac{\Lambda_\text{UV}^2}{m_S^2}\right)c_h\frac{\mu}{M_N^2}SH^\dagger H,
\end{align}
where both $m_{12}$ and $\Lambda_\text{UV}$ denote the UV scale associated with the heavy 2HDM modes $M_H,M_A,M_{H^\pm}$.


\end{document}